\documentclass[12pt,a4paper]{article}
\usepackage[T1]{fontenc}
\usepackage{amsmath,amsfonts,amssymb,amsthm,mathabx}
\usepackage{cite}
\usepackage{hyperref}
\usepackage{graphicx}
\usepackage{stmaryrd}

\newcommand{\be}{\begin{equation}}
\newcommand{\ee}{\end{equation}}
\newcommand{\bea}{\begin{eqnarray}}
\newcommand{\eea}{\end{eqnarray}}

\newcommand{\demi}{\frac{1}{2}}
\newcommand{\phii}{\phi}
\newcommand{\BMS}{\mathrm{BMS}_3}
\newcommand{\bms}{\mathfrak{bms}_3}

\newcommand{\Diff}{\mathrm{Diff}^+(S^1)}

\newcommand{\Vect}{\mathrm{Vect}(S^1)}

\newcommand{\hatVect}{\widehat{\mathrm{Vect}}(S^1)}
\newcommand{\hatDiff}{\widehat{\mathrm{Diff}^+}(S^1)}
\newcommand{\hatbms}{\widehat{\mathfrak{bms}}_3}
\newcommand{\hatBMS}{\widehat{\mathrm{BMS}}_3}

\newcommand{\calG}{\mathfrak{g}}

\newcommand{\calR}{{\cal R}}
\newcommand{\calS}{{\cal S}}
\newcommand{\calT}{{\cal T}}
\newcommand{\calD}{{\cal D}}
\newcommand{\calE}{{\cal E}}
\newcommand{\calF}{{\cal F}}
\newcommand{\calH}{{\cal H}}

\newcommand{\calO}{{\cal O}}

\newcommand{\Ob}{{\cal O}_p}
\newcommand{\lp}{\left(}
\newcommand{\rp}{\right)}
\newcommand{\Ad}{\mathrm{Ad}}
\newcommand{\ad}{\mathrm{ad}}
\newcommand{\SL}{\mathrm{SL}}
\newcommand{\PSL}{\mathrm{PSL}}

\newcommand{\RR}{\mathbb{R}}
\newcommand{\CC}{\mathbb{C}}
\newcommand{\NN}{\mathbb{N}}
\newcommand{\GG}{G\ltimes_{\mathrm{Ad}}\mathfrak{g}_{\text{ab}}}

\def\d_Vphi{\mathrm{d}_V\hspace{-0.06em}\phi}
\def\d_Vphibar{\mathrm{d}_V\hspace{-0.06em}\bar\phi}
\def\d_Vxi{\mathrm{d}_V\hspace{-0.06em}\xi}

\def\be{\begin{eqnarray}}
\def\ee{\end{eqnarray}}
\def\beann{\begin{eqnarray*}}
\def\eeann{\end{eqnarray*}}
\def\beq{\begin{equation}}
\def\eeq{\end{equation}}
\def\ba{\begin{array}}
\def\ea{\end{array}}
\def\ben{\begin{enumerate}}
\def\een{\end{enumerate}}
\def\bea{\begin{eqnarray}}
\def\eea{\end{eqnarray}}

\def\5{\bar }
\def\6{\partial }
\def\7{\hat }
\def\4{\tilde }
\advance\textheight\topskip
\renewcommand{\tilde}{\widetilde}
\renewcommand{\hat}{\widehat}

\renewcommand{\simeq}{\cong}

\newcommand{\dd}{\partial}
\renewcommand{\d}{\partial}

\renewcommand{\geq}{\,{\geqslant}\,}
\renewcommand{\leq}{\,{\leqslant}\,}

\newcommand{\binner}[2]{%
  {\langle}\kern-4.15pt{\langle}#1{,}\,#2{\rangle}\kern-4.15pt{\rangle}}

\newcommand{\half}{\frac{1}{2}}

\newcommand{\ffrac}[2]{\raisebox{.5pt}%
  {\footnotesize$\displaystyle\frac{#1}{#2}$}\kern1pt}

\newcommand{\dover}[2]{\ffrac{\dd #1}{\dd #2}}

\newcommand{\ZZ}{\mathbb{Z}}

\def\cF{\mathcal{F}}

\def\cO{\mathcal{O}}

\numberwithin{equation}{section} \makeatletter
\@addtoreset{equation}{section}

\DeclareFontFamily{OT1}{rsfs}{} \DeclareFontShape{OT1}{rsfs}{m}{n}{
<-7> rsfs5 <7-10> rsfs7 <10-> rsfs10}{}
\DeclareMathAlphabet{\mycal}{OT1}{rsfs}{m}{n}
\def\scri{{\mycal I}}%

\begin{document}

\title{Notes on the BMS group in three dimensions: I. Induced representations}

\author{Glenn Barnich and Blagoje Oblak}

\date{}

\def\mytitle{Notes on the BMS group in three dimensions:\\ I. Induced representations}

\pagestyle{myheadings} \markboth{\textsc{\small G.~Barnich,
    B.~Oblak}}{%
  \textsc{\small BMS$_3$ particles}}

\addtolength{\headsep}{4pt}


\begin{centering}

  \vspace{1cm}

  \textbf{\Large{\mytitle}}


  \vspace{1.5cm}

  {\large Glenn Barnich$^{a}$ and Blagoje Oblak$^{b}$}

\vspace{.5cm}

\begin{minipage}{.9\textwidth}\small \it  \begin{center}
   Physique Th\'eorique et Math\'ematique \\ Universit\'e Libre de
   Bruxelles and International Solvay Institutes \\ Campus
   Plaine C.P. 231, B-1050 Bruxelles, Belgium
 \end{center}
\end{minipage}

\end{centering}

\vspace{1cm}

\begin{center}
  \begin{minipage}{.9\textwidth}
    \textsc{Abstract}. The Bondi-Metzner-Sachs group in three
    dimensions is the symmetry group of asymptotically flat
    three-dimensional spacetimes. It is the semi-direct product of the
    diffeomorphism group of the circle with the space of its adjoint
    representation, embedded as an abelian normal subgroup. The
    structure of the group suggests to study induced representations;
    we show here that they are associated with the well-known
    coadjoint orbits of the Virasoro group and provide explicit
    representations in terms of one-particle states. 
  \end{minipage}
\end{center}

\vfill

\noindent
\mbox{}
\raisebox{-3\baselineskip}{%
  \parbox{\textwidth}{\mbox{}\hrulefill\\[-4pt]}}
{\scriptsize$^a$Research Director of the Fund for Scientific
  Research-FNRS Belgium. E-mail: gbarnich@ulb.ac.be\\
$^b$ Research Fellow of the Fund for Scientific Research-FNRS
Belgium. E-mail: boblak@ulb.ac.be}

\thispagestyle{empty}
\newpage

\begin{small}
{\addtolength{\parskip}{-2pt}
 \tableofcontents}
\end{small}
\thispagestyle{empty}
\newpage

\section{Introduction}
\label{sec:introduction}

\subsubsection*{Motivation}

Asymptotic symmetry groups play an important role in gravitational
theories. From a modern perspective
\cite{tHooft:1993gx,Susskind:1994vu,Maldacena:1997re}, this is because
they are the global symmetry groups for the lower-dimensional, dual
field theories. For example, the asymptotic symmetry group of a
$d$-dimensional anti-de Sitter background is isomorphic to the
conformal group in dimension $d-1$
\cite{Ashtekar:1984,Henneaux:1985tv,Henneaux:1985ey,Ashtekar:1999jx}. From
the point of view of symmetries, $d=3$ is then the most interesting
dimension because the conformal algebra in dimension two is
infinite-dimensional \cite{Brown:1986nw,Belavin:1984vu}.

Historically, however, the first instance of an enhanced,
infinite-dimensional symmetry group in general relativity appeared in
the case of four-dimensional asymptotically flat spacetimes at null
infinity, as shown by Bondi, van der Burg, Metzner and Sachs (BMS)
\cite{Bondi:1962px,Sachs1962a,Sachs1962}. There has recently been a
lot of interest in this group in the context of gravitational
scattering theory and soft graviton theorems
\cite{Strominger:2013jfa,Strominger:2013lka,He:2014laa,Banks:2014iha}
(see also \cite{Duval:2014uva} for a novel approach in the framework
of conformal Carroll groups). In this framework, it is natural to
expect that elementary particles occurring in scattering states should
be classified by representations of the $\mathrm{BMS}_4$ rather than
the Poincar\'e group \cite{Sachs1962,Newman:1965fk,Komar1965,%
  McCarthy1972}.

In line with the point of view taken in two-dimensional conformal
field theories, one can also argue that physics should be constrained
by infinitesimal symmetry transformations that are not necessarily
globally well-defined. When adopting this viewpoint
\cite{Barnich:2009se,Barnich:2010eb,Barnich:2013axa}, the local
symmetry algebra associated with $\mathrm{BMS}_4$ involves two copies
of the Virasoro algebra and an adapted (larger) set of
supertranslations.

Before studying this group head on, it is instructive to first
understand what happens in three dimensions: the $\BMS$ group
\cite{Ashtekar1997} is then again infinite-dimensional. While it is
globally well-defined, its Lie algebra looks roughly like half of that
of the local $\mathrm{BMS}_4$ algebra. More precisely, the
Dirac brackets of its generators form the centrally extended
semi-direct sum of the space of vector fields on the circle with its
adjoint representation, embedded as an abelian ideal
\cite{Barnich:2006avcorr}.

Hence, in an effort to extend some of the successes of holography for
three-dimensional asymptotically anti-de Sitter spacetimes (see
e.g.~\cite{Coussaert:1995zp,Strominger:1998eq,%
  Banados1999,Skenderis:1999nb}) to flat space (see
e.g.~\cite{Barnich:2010eb,Barnich:2011ct,Barnich:2012aw,%
  Barnich:2012xq,Bagchi:2012xr,Barnich:2012rz,Barnich:2013yka}), the
purpose of the present paper is to initiate the study of the
representations of the $\BMS$ group and to show that, like in the
conformal case, they are controlled by powerful, well-established
results on the Virasoro group, which are readily available in the
literature.

More precisely, we will consider induced representations. This is
motivated by the fundamental theorem, due to Mackey (see
e.g.~\cite{Mackey1949,Mackey1952,Mackey1953,MackeyBook,%
A.O.Barut702,Cornwell:1985xs,Cornwell:1985xt}),
stating that -- under suitable regularity assumptions -- the
classification of all unitary irreducible representations of a
semi-direct product group $H=G\ltimes A$, with $A$ abelian, can be
obtained by
\begin{enumerate}

\item determining the set $\hat A$ of characters of $A$;

\item classifying all orbits in $\hat A$ under $G$;

\item computing the stabilizers, the ``little groups'', of the
  orbits;

\item constructing a specific induced representation associated with
  every unitary irreducible representation of a given little group.

\end{enumerate}

The origin of these general group-theoretic results is the celebrated
work by Wigner \cite{Wigner:1939cj} on the classification of the
irreducible representations of the Poincar\'e group.

The $\BMS$ and $\mathrm{BMS}_4$ groups that we are interested in here
have precisely the semi-direct structure required by the theorem. They
fail, however, to satisfy all the regularity assumptions because either
$A$, or both $A$ and $G$, are infinite-dimensional.

The semi-direct product structure of the globally well-defined
$\mathrm{BMS}_4$ group in four dimensions has been clarified in
\cite{Geroch:1971mz,McCarthy1972b}. Its induced representations have
been constructed in \cite{McCarthy1972a,McCarthy1973,McCarthy1973a,%
  McCarthy1975,McCarthy1978,McCarthy1978a}, and shown in
\cite{Piard1977,Piard1977a} to exhaust all
unitary irreducible representations. These results imply, for instance, the
absence of particles with continuous spin.

In three dimensions, the $\BMS$ group is the semi-direct product
$\Diff\ltimes_{\Ad} \Vect_{\text{ab}}$ of the group of
orientation-preserving diffeomorphisms of the circle with its Lie
algebra embedded as an abelian normal subgroup, so that both factors
are infinite-dimensional. The property that makes this semi-direct
product special, and is shared by the Poincar\'e group in three
dimensions -- a subgroup of $\BMS$ --, is that the non-abelian group
$G$ acts on the abelian group $A$ through the adjoint action,
$\GG$. In this case, the orbits relevant for induced representations
coincide with the coadjoint orbits of $G$. The latter are endowed with
a canonical symplectic structure and, upon quantization, play an
important role in the representation theory of $G$ itself (see
e.g.~\cite{A.A.Kirillov897}). The main difference here is that, for
representations of $\GG$, there is no need to quantize the coadjoint
orbits. Their symplectic structure does nevertheless play
an important role, as it provides an invariant measure needed for
induced representations.

In the case of $\BMS$ and its central extension $\hatBMS$, induced
representations are thus classified by coadjoint orbits of
$\Diff$ and of the Virasoro group $\hatDiff$, which have been extensively
studied in the literature (see e.g.~\cite{LazPan,Segal:1981ap} and
also \cite{Witten:1987ty,Bakas:1988mq,Taylor:1992xt} for
discussions in the physics literature).

The $\bms$ algebra has also appeared in the framework of two-dimensional
statistical systems in the form of an infinite-dimensional extension
of $\mathfrak{alt}_1\simeq \mathfrak{iso}(2,1)$ \cite{Henkel2005}. It
is also isomorphic to the infinite-dimensional extension of the
Galilean conformal algebra in two dimensions, $\mathfrak{gca}_2$
\cite{Bagchi:2009my,Hosseiny:2009jj,Bagchi:2009pe}. The classification
of induced representations of $\BMS$ is thus also applicable in these
cases, and governs the physics of these problems.

A companion paper will be devoted to the coadjoint representation of
$\hatBMS$, which coincides with the reduced phase space of
three-dimensional asymptotically flat gravity. In particular, we will
establish a direct connection between induced representations and the
quantization of coadjoint orbits of $\hatBMS$.

\subsubsection*{Plan of the paper}

The paper is organized as follows. In section \ref{sec:induc-repr}, we
briefly review relevant results on induced representations. This is
followed by remarks on projective representations \cite{Bargmann1954}
(see e.g.~\cite{Simms1971,Weinberg:1995mt} for a summary). In section
\ref{sec:repr-poinc-group}, we then apply the inducing construction to
the example of the Poincar\'e group in three dimensions
\cite{Binegar1982}.

In section \ref{sec:infin-dimens-aspects}, we collect some remarks on
the Virasoro algebra and group, following in particular
\cite{Ovsienko2005,guieu2007,Khesin2009} (see also
\cite{Unterberger:2011yya} for a detailed analysis of a group with a
structure similar to that of $\BMS$). These preliminaries are then used in
section \ref{sec:bms3-group} to discuss the structure of the $\BMS$
group and of its extensions. We also review the realization of $\BMS$
as a transformation group on null infinity $\scri$, and isolate its
Poincar\'e subgroups.

In section \ref{sec:induc-repr-bms3} on induced representations of
$\BMS$, we start by reviewing the Virasoro coadjoint orbits. We also
discuss energy bounds, mostly following \cite{Balog:1997zz} to which
we refer for a more complete and self-contained review (see also
\cite{guieu2007}). We then describe features of $\BMS$ particles, that
is, induced representations of $\hatBMS$, by taking inspiration from
the Poincar\'e group in three dimensions.

Finally, section \ref{sec:issues-questions} is devoted to an
enumeration of open issues, some of which we hope to address
elsewhere.

\section{Induced representations of semi-direct products}
\label{sec:induc-repr}

In this section we review the construction of induced
representations for separable, locally compact topological
groups. In particular, here and below ``topological group'' will mean
that all these regularity assumptions are verified. It will also be
understood that all group homomorphisms, actions and representations
are continuous. We end the section with a brief review on projective
representations.

\subsection{Semi-direct products}

Let $A$ and $G$ be two topological groups; elements of $G$ are denoted
as $f$, $g$, etc. and elements of $A$ as $\alpha$, $\beta$, etc. For
our purposes, $A$ is assumed to be an abelian additive vector group
throughout, and therefore the group operation in $A$ is denoted as the
addition $+$. Let $\mathrm{GL}\lp A\rp$ be the linear group of $A$ and
let $\sigma:G\rightarrow\mathrm{GL}\lp A\rp, f\mapsto\sigma_f$ be a
group homomorphism. The {\it semi-direct product} of $G$ and $A$ is
the group
\begin{equation*}
H:= G\ltimes_{\sigma}A,
\end{equation*}
whose elements are
pairs $\lp f,\alpha\rp$, with a group operation given by 
\begin{equation}
\label{semiGroupLaw}
\lp f,\alpha\rp\lp g,\beta\rp:=\lp fg,\alpha+\sigma_f\beta\rp.
\end{equation}
Typical examples of groups of this form include the euclidean groups
$\mathrm{SO}(n)\ltimes\RR^n$ and the Poincar\'e groups
$\mathrm{SO}(n-1,1)^{\uparrow}\ltimes\RR^n$. 

If $G$ is a Lie group, let $\calG$ be the corresponding Lie algebra,
whose elements are written as $X$, $Y$, etc.  Since $A$ is a vector
group, it is isomorphic -- as a vector space -- to its Lie algebra,
which is why we continue to denote elements of the latter by $\alpha$,
$\beta$, etc. If $\mathrm{End}\lp A\rp$ is the commutator algebra of
linear operators in $A$, the differential of the map $\sigma$ at the
origin is the Lie algebra homomorphism
$\Sigma:\calG\rightarrow\mathrm{End}\lp A\rp,X\mapsto\Sigma_X$.  Then
the Lie algebra of the semi-direct product $H$ is the {\it semi-direct
  sum} $\mathfrak{h}=\calG\oright_{\Sigma} A$. Writing elements of the
latter as pairs $(X,\alpha)$, the Lie bracket in $\mathfrak{h}$ reads
\begin{equation}
\label{triangle}
\left[\lp X,\alpha\rp,\lp
  Y,\beta\rp\right]=\lp\left[X,Y\right],\Sigma_X\beta -\Sigma_Y\alpha\rp.
\end{equation}

\subsection{Orbits and little groups}
\label{subsec-orbits-littleGroups}

Since $A$ is abelian, its unitary irreducible representations are
necessarily one-dimensional. As a consequence, they are fully
determined by their character, which can be written as
\begin{equation*}
\chi:A\rightarrow\CC,\ \alpha\mapsto\chi(\alpha)=e^{i\langle p,\alpha\rangle},
\end{equation*}
where the map $\langle p,.\rangle:A\rightarrow\RR$ is a continuous
homomorphism, assumed to be smooth in the case of Lie groups. The map
$p$ belongs to $A^*$, the topological dual of $A$, which is thus in
one-to-one correspondence with the set of inequivalent unitary
irreducible representations of $A$. Elements of $A^*$ will be denoted
by $p$, $q$, etc.

The action of $G$ on $A^*$ is the homomorphism
$\sigma^*:G\rightarrow\mathrm{GL}\lp A^*\rp$,
$f\mapsto\sigma^*_f$ given by
\begin{equation}
\label{sigma*}
\langle \sigma^*_fp,\alpha\rangle
:=
\langle p,\sigma_{f^{-1}}\alpha\rangle.
\end{equation}
For simplicity, we will use the notation $f\cdot p:=\sigma^*_fp$
below. The {\it orbit} of some $p\in A^*$ is then defined as the set
\begin{equation}
\calO_p:=\left\{f\cdot p|f\in G\right\}\subset A^*.\label{eq:30}
\end{equation}
The {\it little group} $G_p$ for this orbit is the stationary subgroup
of $G$ for the element $p$ and the action $\sigma^*$,
\begin{equation}
G_p:=\left\{f\in G|f\cdot p=p\right\}.\label{eq:31}
\end{equation}
By construction, the orbit ${\cal O}_p$ is a homogeneous space for the
action $\sigma^*$ and, under suitable regularity assumptions, ${\cal
  O}_p$ is homeomorphic to the (left) coset space $G/G_p$. This
identification can be made explicit by introducing a map
\begin{equation}
\label{mapg}
g:{\cal O}_p\rightarrow G, q\mapsto g_q,
\end{equation}
where $g_q$ is such that
$g_q\cdot p=q$ for all $q\in{\cal O}_p$. One may then identify
$q\in{\cal O}_p$ with the left coset $g_qG_p\in G/G_p$. The map $g$ is
defined up to multiplication from the right by any map which sends
$\Ob$ on $G_p$.

\subsection{Induced representations}
\label{sec:inducedRepsBla}

\subsubsection*{Generalities}

Consider a unitary representation
$\calR:G_p\rightarrow\mathrm{GL}\lp\calE\rp:f\mapsto\calR[f]$ of a
given little group $G_p$ in a separable Hilbert space $\calE$,
equipped with a scalar product $(.|.)$. A representation of the
subgroup $H_p:= G_p\ltimes_{\sigma}A$ of $H$, where it is understood
that $\sigma$ is restricted to $G_p$, is given by
\begin{equation}
  \calS\left[\lp f,\alpha\rp\right]:=
  e^{i\langle p,\alpha\rangle}\,\calR[f]\quad\forall\,\lp f,\alpha\rp\in H_p.\label{eq:33}
\end{equation}
Let $\mu$ denote a $G$-invariant measure on the orbit $\Ob$ and
consider the Hilbert space ${\cal H}$ of maps
$\Psi:\Ob\rightarrow\calE$ that are square-integrable with respect to
this measure, the scalar product in $\calH$ being defined as
\begin{equation}
\langle\Phi|\Psi\rangle:=\int_{\Ob}d\mu(q)\left(\Phi(q)|\Psi(q)\right).
\label{scalarProd}
\end{equation}

The representation $\calT$ {\it induced} by $\calS$ is then defined \cite{A.O.Barut702,Cornwell:1985xs} as the
homomorphism $\calT:H=G\ltimes_{\sigma}A\rightarrow\mathrm{GL}\lp{\cal
  H}\rp, (f,\alpha)\mapsto\calT\left[\lp f,\alpha\rp\right]$ with 
\begin{equation}
\label{calt}
\Big(\calT\left[\lp f,\alpha\rp\right]\Psi\Big)(q) :=
e^{i\langle q,\alpha\rangle}\,\calR\left[g_q^{-1}fg_{f^{-1}\cdot q}\right]\Psi\lp
f^{-1}\cdot q\rp
\quad\forall\,
q\in\Ob, \forall\,\Psi\in {\cal H}.
\end{equation}
One easily verifies that this representation is unitary with respect
to the scalar product (\ref{scalarProd}). Note that
$g_q^{-1}fg_{f^{-1}\cdot q}$ belongs to the little group $G_p$ for all
$f\in G$ and any $q\in\Ob$.

As briefly recalled in the introduction, Mackey's main theorem
\cite{Mackey1949,Mackey1952,Mackey1953,MackeyBook,A.O.Barut702} for
separable, locally compact groups states that, provided the
semi-direct product $H$ is regular, {\it every} unitary irreducible
representation of $H$ is unitarily equivalent to an induced
representation.

\subsubsection*{Delta function basis}

Definition (\ref{calt}) of the induced representation $\calT$ can be written in
an alternative form, that can sometimes be more convenient
\cite{Cornwell:1985xs}. Consider the $G$-invariant delta function
$\delta_{\mu}$ associated with the measure $\mu$ on $\calO_p$,
\begin{equation*}
  \int_{\calO_p}d\mu(q)\delta_{\mu}(q-k)\varphi(q)=\varphi(k) 
\quad \forall\varphi,\;\forall\,k\in\calO_p.
\end{equation*}
Then, if $\{e_m|m=1,2,3,\dots\}$ is an orthonormal basis of $\cal E$,
define the states
\begin{equation}
\Psi_{k,m}(q):=\delta_{\mu}(q-k)e_m,\label{deltaBasis}
\end{equation}
which form an (improper) orthonormal basis of ${\cal H}$:
\begin{equation}
\left<\Psi_{k,m}|\Psi_{\ell,n}\right> =
\delta_{mn}\delta_{\mu}(k-\ell).
\label{scalPlane}
\end{equation}
The induced representation (\ref{calt}) then acts on these states
according to
\begin{equation}
  \label{eq:1}
  \calT\left[\lp f,\alpha\rp\right]\Psi_{k,m} =
  e^{i\langle {f}\cdot k,\alpha\rangle}{{\lp\calR\left[g_{f\cdot
k}^{-1}\,f\,g_{k}\right]\rp}^{n}}_{m}\Psi_{{f}\cdot k,n}.
\end{equation}
Note that, assuming irreducibility of $\calR$, the irreducibility of
$\calT$ is obvious in this formulation: since $\calR$ is irreducible,
and since the delta function basis is defined only on the orbit
$\calO_p$, there always exists a transformation $f$ that maps
$\Psi_{k,m}$ onto $\Psi_{\ell,n}$, for all $k,\ell$ and any $m,n$.

\subsection{The case of the adjoint action}
\label{sec:adjo-repr}

A particular case, that will be relevant for $\BMS$, is the case where
$G$ is a Lie group, $A=\mathfrak g_{\text{ab}}$ is its Lie algebra (seen as the
abelian additive group of a vector space) and $\sigma=\Ad$ is the
adjoint action, 
\begin{equation}
  H=\GG.
  \label{GG}
\end{equation}
In this case, the orbits $\Ob$ defined in (\ref{eq:30}) coincide with
the coadjoint orbits of $G$. The little groups $G_p$ are Lie groups
and the associated Lie algebras, the little algebras, are denoted by
$\mathfrak g_p$.

The Lie algebra of (\ref{GG}) is the semi-direct sum 
\begin{equation}
\label{algebrah}
\mathfrak{h}=\calG\oright_{\Sigma}\calG_{\text{ab}},
\end{equation}
where $\calG_{\text{ab}}$ denotes the {\it abelian} Lie algebra that
is isomorphic to $\calG$ as a vector space. According to
(\ref{triangle}), the Lie bracket in $\mathfrak{h}$ then reads 
\begin{equation}
\left[\lp X,\alpha \rp,\lp Y,\beta\rp\right]=
\lp\left[X,Y\right],[X,\beta]-[Y,\alpha]\rp.\label{eq:74}
\end{equation}

The additional structure that is available in this setting is a
Poisson bracket on $\mathfrak{g^*}$ \cite{A.A.Kirillov897}. In the
finite-dimensional case, if $x_a$ are coordinates on $\mathfrak g^*$
associated with a dual basis $e^a_*$, the bracket is given by
\begin{equation*}
  \{F,G\}=\dover{F}{x_a} C^c_{ab}\dover{G}{x_b} x_c, \quad F,G\in
  C^\infty(\mathfrak g^*),
\end{equation*}
where $C^a_{bc}$ are the structure constants in the basis $e_a$ of
$\mathfrak g$. If $G$ is connected, its coadjoint orbits are the
leaves of the symplectic foliation induced by this bracket on
$\mathfrak g^*$. The coadjoint action of a group element $f$ preserves
these leaves and is generated by a suitable
Hamiltonian. The $G$-invariant symplectic form $\omega$ on $\calO_p$
is then defined by
\begin{equation}
\omega_q\left(\ad^*_Xq,\ad^*_Yq\right)
:=
\big<q,[X,Y]\big>
\quad\forall\,q\in\calO_p,\,\forall\,X,Y\in\mathfrak{g}.
\label{KK}
\end{equation}
Note that the tangent space of $\calO_p$ at $q$ consists of coadjoint
vectors of the form $\ad^*_Xq$. If the orbit $\calO_p$ is
$2n$-dimensional, the $2n$-form $\omega^n/n!$ is a $G$-invariant
volume form on $\calO_p$; when considering induced representations of
$\GG$, it can be used to define the scalar product (\ref{scalarProd}).

\subsection{Remarks on projective representations}
\label{sec:remarks-proj-repr}

In applications to quantum mechanics, one is primarily interested in
projective unitary representations, i.e. unitary representations up to
a phase \cite{Weinberg:1995mt}. Bargmann's results \cite{Bargmann1954}
state that each projective representation of a group $G$ can be
identified with an ordinary representation of the central extension
$\hat G$ of $G$. The latter is determined by the second cohomology
group $H^2(G,\RR)$.

One can show that for a connected Lie group $G$ with vanishing Lie
algebra cohomology in degree two, projective representations are
standard representations multiplied by phase factors forming a
representation of the fundamental group $\pi_1(G)$ of $G$. If in
addition $G$ is simply connected, then $H^2(G,\RR)=0$ and every
projective unitary representation of $G$ can be lifted to an ordinary
unitary representation.

This is the reason why, in discussing representations, one considers
the universal covering $\SL(2,\CC)\ltimes L^2(S^2)$ of the globally
well-defined BMS$_4$ group, and also the centrally extended Virasoro
group $\hatDiff$ and its universal cover.

\section{The Poincar\'e group in three dimensions}
\label{sec:repr-poinc-group}

In this section, we briefly recall results on representations of the
Poincar\'e group in three dimensions, the exact isometry group of
three-dimensional Minkowski spacetime, in order to compare them with
those of the BMS$_3$ group. This is motivated by the fact that (i) the
Poincar\'e group is a subgroup of BMS$_3$, (ii) it satisfies all the
assumptions required by Mackey's theorem on induced representations,
and (iii) it is a simple example of a semi-direct product of the form
$\GG$.

\subsection{Double cover of the Poincar\'e group}
\label{sec:double-cover}

The Poincar\'{e} group in three spacetime dimensions is the
semi-direct product of the connected Lorentz group
$\mathrm{SO}\lp2,1\rp^{\uparrow}$ with the abelian vector group
$\RR^3$ of spacetime translations:
\begin{equation*}
P^\uparrow_3:=\mathrm{SO}\lp2,1\rp^{\uparrow}\ltimes_{\sigma}\RR^3,
\label{eq:35}
\end{equation*}
where $\sigma$ is the vector representation of
$\mathrm{SO}(2,1)$. Explicitly, if elements of the Poincar\'{e} group
are written as pairs $\lp\Lambda,a\rp$, where
$\Lambda\in\mathrm{SO}\lp2,1\rp^{\uparrow}$ and
$a=(a^0,a^1,a^2)\in\RR^3$, then Lorentz transformations act on
translations according to $\lp\sigma_{\Lambda}
a\rp^\mu=\Lambda^{\mu}_{\;\;\nu}a^{\nu}$ with
$\mu,\nu,\ldots=0,1,2$. 

When one is interested in projective representations, it is useful to
consider, instead of $P^\uparrow_3$, its double cover 
\begin{equation}
\overline{P}^\uparrow_3:=
\SL\lp2,\RR\rp\ltimes_{\Ad}\mathfrak{sl}\lp2,\RR\rp_{\text{ab}},\label{eq:8}
\end{equation}
which follows from the standard isomorphism
$\mathrm{SO}\lp2,1\rp^{\uparrow}\cong\mathrm{SL}\lp2,\RR\rp/\mathbb{Z}_2
:=\mathrm{PSL}(2,\RR)$. To show this isomorphism, one begins by
identifying $\RR^3$ with $\mathfrak{sl}(2,\RR)_{\text{ab}}$,
associating with a translation vector $a$ a traceless $2\times 2$
matrix $\alpha:= a^{\mu}t_{\mu}$, where the generators $t_{\mu}$ of
$\mathfrak{sl}\lp2,\RR\rp$ can be chosen as
\begin{equation}
\label{generators}
t_0=\frac{1}{2}
\begin{pmatrix}
0 & 1\\
-1 & 0
\end{pmatrix},
\quad
t_1=\frac{1}{2}
\begin{pmatrix}
0 & 1\\
1 & 0
\end{pmatrix},
\quad
t_2=\frac{1}{2}
\begin{pmatrix}
1 & 0\\
0 & -1
\end{pmatrix}.
\end{equation}
They satisfy $t_\mu t_\nu =\half
\epsilon_{\mu\nu\rho}t^\rho+\frac{1}{4}\eta_{\mu\nu}\mathbb{I}$, where
$\eta_{\mu\nu}:={\rm diag}(-1,1,1)$ and $\epsilon_{012}:=1$; as usual,
$\eta_{\mu\nu}$ and its inverse will be used below to lower and raise
indices. The invariant bilinear form on
$\mathfrak{sl}\lp2,\RR\rp$ reproduces the Minkowskian scalar product
of vectors:
\begin{equation}
\label{invQuad}
\langle\alpha,\beta\rangle\:= 2\mathrm{Tr}\left(\alpha\beta\right)=
\eta_{\mu\nu}a^{\mu}b^{\nu}.
\end{equation}
The isomorphism then follows from the surjective
homomorphism
$H:\SL\lp2,\RR\rp\rightarrow\mathrm{SO}\lp2,1\rp^\uparrow$,
$f\mapsto H[f]$ such that 
\begin{equation}
f t_{\mu} f^{-1} a^{\mu}= t_\nu
\left( H [f] \right)^{\nu}_{\;\;\rho}a^{\rho} \quad
\forall\,a^{\mu}t_{\mu}\in\mathfrak{sl}\lp2,\RR\rp,\label{eq:43}
\end{equation}
whose kernel is isomorphic to $\ZZ_2$. For later use, we note that if 
\begin{equation*}
  f=\begin{pmatrix}
a & b \\ c & d
\end{pmatrix}\in \SL\lp2,\RR\rp,
\end{equation*}
then 
\begin{equation}
\label{Phi}
 H[f]=
\begin{pmatrix}
\demi\lp a^2+b^2+c^2+d^2\rp & \demi\lp a^2-b^2+c^2-d^2\rp & -ab-cd\\
\demi\lp a^2+b^2-c^2-d^2\rp & \demi\lp a^2-b^2-c^2+d^2\rp & -ab+cd\\
-ac-bd & bd-ac & ad+bc
\end{pmatrix}.
\end{equation}
From now on, when talking about
the Poincar\'e group in three dimensions, we will refer to its double
cover (\ref{eq:8}).

It follows from the structure $\GG$ of the Poincar\'e group that its
Lie algebra has the general form (\ref{algebrah}). 
In terms of the basis 
\begin{equation*}
  j_{-1}:=-i(t_0+t_1),\quad j_1:=-i(t_0-t_1),\quad j_0:=-i t_2
\end{equation*}
of $\mathfrak{sl}(2,\RR)$ and of its counterpart
$\left\{p_{-1},p_1,p_0\right\}$ in $\mathfrak{sl}(2,\RR)_{\text{ab}}$,
the commutation relations of the three-dimensional Poincar\'e algebra
are 
\begin{equation}
i\left[j_m,j_n\right] = (m-n)j_{m+n},\quad i\left[j_m,p_n\right] =
(m-n)p_{m+n},\quad i\left[p_m,p_n\right] = 0,\label{eq:6}
\end{equation}
with $m,n=-1,0,1$. The $j_m$'s are thus to be interpreted as Lorentz
generators, while the $p_m$'s are generators of translations.

Note that $\SL\lp2,\RR\rp$ is not simply connected: its fundamental
group is isomorphic to $\ZZ$, as is that of the Poincar\'e group in
three dimensions. The latter thus admits genuinely projective unitary
representations, whose classification relies on its universal cover.

\subsection{Induced representations of the Poincar\'e group}
\label{sec:orbits-little-groups}

\subsubsection*{Coadjoint orbits of $\SL(2,\RR)$}

The existence of the invariant bilinear form \eqref{invQuad} implies
that adjoint and coadjoint representations of $\SL(2,\RR)$ are
equivalent, so that coadjoint orbits can be deduced from
adjoint ones. If $\{t^{0*},t^{1*},t^{2*}\}$ denotes the dual basis of
$\mathfrak{sl}(2,\RR)^*$ associated with (\ref{generators}), write
$p=p_\mu t^{\mu*}\in\mathfrak{sl}\lp2,\RR\rp^*$. The equivalence
follows by associating, with each such coadjoint vector $p$, the
translation vector $p^{\mu}t_{\mu}\in\mathfrak{sl}(2,\RR)$.

Fix an arbitrary number $\kappa >0$. It can then be shown that each
coadjoint orbit $\Ob$ of $\SL(2,\RR)$ is of one of the following six
types:

\begin{enumerate}

\item Upper hyperboloid $q_\mu q^\mu =-m^2$, $q_0>0$, $m>0$. Orbit
  representative $p=mt^{0*}$. Little group $\mathrm{U}(1)$.

\item Lower hyperboloid $q_\mu q^\mu =-m^2$, $q_0<0$, $m>0$. Orbit
  representative $p=-mt^{0*}$. Little group $\mathrm{U}(1)$.

\item Upper null cone $q_{\mu}q^{\mu}=0$, $q_0>0$. Orbit
  representative $p=\kappa(t^{0*}+t^{1*})$. Little group
  $\RR\times\ZZ_2$.

\item Lower null cone $q_{\mu}q^{\mu}=0$, $q_0<0$. Orbit
  representative $p=-\kappa(t^{0*}+t^{1*})$. Little group
  $\RR\times\ZZ_2$.

\item One-sheeted hyperboloid $q_{\mu}q^{\mu}=m^2$. Orbit
  representative $p=mt^{2*}$. Little group $\RR\times\ZZ_2$.

\item Trivial orbit $p=0$. Little group $\SL(2,\RR)$.
  
\end{enumerate}

\subsubsection*{Energy-momentum}

When interpreting unitary irreducible representations of the
Poincar\'e group as {\it one-particle states} of a relativistic
particle, $p$ is its energy-momentum vector. Particles on the
lower/upper hyperboloids are massive, the sign of $p_0$ determining
whether they propagate towards the future or the past, while particles
on the one-sheeted hyperboloids move faster than light -- they
are tachyons.

\subsubsection*{Spin}

In building induced representations, one also needs unitary
irreducible representations of the little groups. In the massive case
with ${\rm U}(1)$ little groups, these representations are
one-dimensional, and are labelled by an integer $j$. When considering
projective representations, the universal cover of the Poincar\'e
group becomes relevant and these ${\rm U}(1)$'s are replaced by $\RR$.
The label $j$ becomes an arbitrary real number and there is no
quantization of spin.

In the massless case, the unitary irreducible representations of the
little groups $\RR\times\ZZ_2$ are characterized by spins, also called
helicities, $s\in \RR$ and $\epsilon=\pm 1$.

Finally, the trivial orbit has little group $\SL(2,\RR)$, which admits
three series of infinite-dimensional unitary irreducible
representations. The trivial representation corresponds to the vacuum
state.

\subsubsection*{Interpretation of induced representations}

The transformations $g_q$ introduced in (\ref{mapg}), and defined so
that $g_q\cdot p=q$, correspond here to standard boosts that map an
orbit representative $p$ (e.g. the momentum of a massive particle at
rest) on a momentum $q$ belonging to its orbit (representing the
momentum of the same particle with non-zero velocity). The
composite Lorentz transformation $g_q^{-1}\,f\,g_{f^{-1}\cdot q}$
appearing in (\ref{calt}) is then called a {\it Wigner rotation}.

For the orbits $\calO_p$ of future-directed massive particles in three
dimensions, the Lorentz-invariant measure appearing in the scalar
product (\ref{scalarProd}) is given by the volume form
\begin{equation}
\label{LorentzMeas}
d\mu(q)=\frac{dq_1dq_2}{2q_0},\quad q_0=\sqrt{m^2+q_1^2+q_2^2}.
\end{equation}
This form coincides (up to normalization) with the
symplectic form (\ref{KK}) on $\calO_p$.

For all orbits except the trivial one, the little group is
one-dimensional and the corresponding space $\calE$ introduced above
(\ref{eq:33}) is just $\CC$, so that the Hilbert space $\calH$ is the
space of complex-valued quantum wavefunctions in momentum space. The
induced representation (\ref{calt}) acts by simultaneously boosting,
rotating and translating these wavefunctions. Each element $\Psi_k$ of
the delta function basis (\ref{deltaBasis}) then represents a plane
wave of definite momentum $k$ and spin $j$.

\section{Virasoro algebra and group}
\label{sec:infin-dimens-aspects}

Here we recall some basic definitions regarding the Virasoro group,
expressed in terms of functions on the circle. We refer for instance
to \cite{guieu2007} for a more detailed review. 

\subsection{Virasoro algebra}
\label{sec:virasoro-algebra}

\subsubsection*{Vector fields on the circle}

Smooth functions on $S^1$ can be identified with $2\pi$-periodic
smooth functions of a variable $\phii\in \RR$. Let $\Vect$ denote the
Lie algebra of vector fields on the circle. Its elements are written
as $X=X(\phii)\frac{\partial}{\partial\phii}$, where $X(\phii)$ is a
smooth function, the Lie bracket being
$[X,Y]=(XY'-YX')\dover{}{\phii}$. In terms of the Fourier basis
$\ell_m:= i e^{im\phii}\dover{}{\phii}$, the commutation relations of
$\Vect$ read
\begin{equation*}
[\ell_m,\ell_n]=(m-n)\ell_{m+n}.
\end{equation*}

\subsubsection*{Tensor densities on the circle}

The space of tensor densities $\cF_\lambda$ of degree $\lambda\in \RR$
consists of elements of the form
$\alpha=\alpha(\phii)d\phii^{\lambda}$, with $\alpha(\phii)$ a smooth
function. It is a $\Vect$-module with action 
\begin{equation}
  \label{eq:15a}
  \Sigma^{\lambda}_X\alpha= X\alpha'+\lambda X'\alpha.
\end{equation}
In particular, $\cF_{-1}\simeq\Vect$ as a
$\Vect$-module, and the action of
$\Vect$ on $\cF_{-1}$ coincides with its adjoint action. 

Note that the bilinear form on $\cF_\lambda\times\cF_{1-\lambda}$ given by
\begin{equation}
  \label{eq:bilin}
\langle\alpha,\beta\rangle:=
\int_{S^1}\alpha\otimes\beta\quad\text{for}\;\alpha\in\cF_{\lambda},
\;\beta\in\cF_{1-\lambda} 
\end{equation}
is invariant under the action of $\Vect$ in the sense that $\langle
\Sigma^{\lambda}_X\alpha,\beta\rangle+\langle\alpha,\Sigma^{1-\lambda}_X\beta\rangle=0$. This
bilinear form can be used to identify the regular dual $\Vect^*$ with
the space $\cF_{2}$ of quadratic densities: for $p=p(\phii)d\phii^2\in
\cF_{2}$,
\begin{equation*}
\langle p,\alpha\rangle
=\int_{0}^{2\pi}\,p(\phii)\alpha(\phii)
\,d\phii\quad\forall\;\alpha\in\calF_{-1}. 
\end{equation*}
The action of $\Vect$ on $\cF_2$ then coincides with its coadjoint
action: $\Sigma^2_Xp=\ad^*_Xp=(Xp'+2 X' p)d\phii^2$.

\subsubsection*{Centrally extended algebra}

The cohomology ring $H^*(\Vect,\RR)$ is
generated by elements in degrees $0,2,3$, where the cohomology group is
one-dimensional. In particular, a representative in degree $2$ is
given by the {\it Gelfand-Fuchs cocycle}
\begin{equation}
  C(X,Y)= \int^{2\pi}_{0}d\phii\ {\rm det}
\begin{pmatrix} X' & Y' \\
    X'' & Y''\end{pmatrix}, \label{eq:11}
\end{equation}
whose expression in the Fourier basis is
$C\left(\ell_m,\ell_n\right)=4\pi im^3\delta^0_{m+n}$. The second
cohomology group $H^2(\Vect,\RR)$ is directly related to
$H^1(\Vect,\cF_2)$, which is also one-dimensional, with representative
$s[X] =2 X''' d\phii^2$.

The Virasoro algebra $\hatVect$ is the universal central extension of
$\Vect$ and consists of pairs $(X,-ia)$, where $a\in \RR$. The
associated Lie bracket reads
\begin{equation}
\label{viralg}
  \left[(X,-ia),(Y,-ib)\right]
  =\Big([X,Y],-\frac{i}{48\pi}C(X,Y)\Big).
\end{equation}
Equivalently, in terms of the Fourier basis
$L_m:=(\ell_m,\frac{1}{24}\delta_m^0)$, $Z:=(0,1)$, the commutation
relations of the Virasoro algebra read
\begin{equation*}
  \left[L_m,L_n\right]=(m-n)L_{m+n}+\frac{1}{12}
\delta^0_{m+n}m(m^2-1) Z,\quad [Z,L_m]=0.
\end{equation*}
The dual space $\hatVect^*$ of the Virasoro algebra consists of pairs
$(p,ic)$ with $c\in\RR$, paired with $\hatVect$ according to
\begin{equation}
  \label{eq:18}
  \langle (p,ic),(X,-ia)\rangle=\int_0^{2\pi}d\phii\, pX\, +ca.
\end{equation}
The coadjoint representation of the Virasoro algebra then reads
\begin{equation}
  \label{eq:17}
  \ad^*_{X} (p,ic)=
  \left(\left[Xp'+2X'p-\frac{c}{24\pi}X'''\right] d\phii^2,0\right).
\end{equation}
Here and below, we drop the central element $-ia$ in the subscript of
the coadjoint action, since it acts trivially.

\subsection{Virasoro group}
\label{sec:virasoro-group}

\subsubsection*{Diffeomorphisms of the circle}

The group of orientation-preserving diffeomorphisms of the circle will
be denoted by $\Diff$. It can be endowed with the structure of a
Lie-Fr\'echet group with Lie algebra $\Vect$. Similar notations and
properties hold when replacing $S^1$ with $\RR$. $\Diff$ is
connected, but not simply connected: its fundamental group is
$\pi_1(\Diff)\simeq 
\mathbb Z$. Its universal cover $\widetilde{\mathrm{Diff}^+}(S^1)$,
whose elements will be denoted as $f$, $g$, etc., is the subgroup of
$\mathrm{Diff}^+(\RR)$ consisting of $2\pi\ZZ$-equivariant
diffeomorphisms
\begin{equation}
f:\RR\rightarrow\RR,\phii\mapsto f(\phii)\quad\text{such that}\;
f'(\phii)>0\;\text{and}\;f(\phii+2\pi)=f(\phii)+2\pi.\label{eq:56}
\end{equation}
The corresponding projection
$\chi:\widetilde{\mathrm{Diff}^+}(S^1)\rightarrow \Diff,\,f\mapsto F$
is defined through
\begin{equation}
  \label{eq:56a}
  e^{i f(\phii)}=F(e^{i\phii}),
\end{equation}
so that
\begin{equation*}
  \Diff=\widetilde{\mathrm{Diff}^+}(S^1)/2\pi \ZZ, 
\end{equation*}
where $2\pi\ZZ$ is identified with the subgroup of
$\widetilde{\mathrm{Diff}^+}(S^1)$ consisting of translations of $\RR$
by integer multiples of $2\pi$.

For future use, it is convenient to associate with
$f\in\widetilde{\mathrm{Diff}^+}(S^1)$ the function
\begin{equation}
\theta(\phi):=\frac{1}{(f^{-1})'(\phi)},\label{eq:4}
\end{equation}
which is positive, $2\pi$-periodic, and satisfies
\begin{equation}
  \label{eq:2}
  \int_0^{2\pi}\frac{d\phii}{\theta(\phi)}=2\pi.
\end{equation}

\subsubsection*{Tensor densities on the circle}

$\widetilde{\mathrm{Diff}^+}(S^1)$ acts on $\cF_\lambda$ according to 
\begin{equation}
  \label{eq:17a}
  \sigma^{\lambda}_{f^{-1}}\lp\alpha\rp
  := (f')^{\lambda}\,\alpha\circ f.
\end{equation}
For $\lambda=-1$, this action coincides with the adjoint action when
taking into account that $\cF_{-1}\simeq\Vect$. The bilinear form
defined in (\ref{eq:bilin}) is invariant in the sense that $\langle
\sigma^{\lambda}_f\alpha,\sigma^{1-\lambda}_f\beta\rangle
=\langle\alpha,\beta\rangle$.

\subsubsection*{Centrally extended group}

The second cohomology group
$H^2(\widetilde{\mathrm{Diff}^+}(S^1),\RR)$ is one-dimensional and can
be represented by the {\it Bott-Thurston cocycle}
\begin{equation}
  \label{eq:15}
  B(f,g)=
  \int_{0}^{2\pi}d\phii\,\ln(f'\circ
  g)\left(\ln\left(g'\right)\right)'. 
\end{equation}

For simplicity, we define\footnote{There is an alternative definition
  of both the Virasoro and the $\BMS$ groups that involves $\Diff$
  instead of its universal cover. What we call the
  Virasoro and $\BMS$ groups here corresponds to the universal cover
  of these groups in the alternative description.} the {\it Virasoro
  group} $\hatDiff=\widetilde{\mathrm{Diff}^+}(S^1)\times \RR$ to be
the central extension of $\widetilde{\mathrm{Diff}^+}(S^1)$. It
consists of pairs $(f,-ia)$, the group law being
\begin{equation*}
  (f,-ia)\cdot(g,-ib)=\left(f\circ g,-i\Big(a+b
-\frac{1}{48\pi}B(f,g)\Big)\right).
\end{equation*}
The Lie algebra of the Virasoro group is the Virasoro algebra
(\ref{viralg}), up to an overall sign in the commutation relations.

As in the case of $\Vect$, there is a direct relation between
$H^2(\widetilde{\mathrm{Diff}^+}(S^1),\RR)$ and
$H^1(\widetilde{\mathrm{Diff}^+}(S^1),\cF_{2})$. The cocycle
associated in this way with $B(f,g)$ is $2S[f]d\phii^2$, where 
\begin{equation*}
  S[f]:=\frac{f'''}{f'}-\frac{3}{2}\left(\frac{f''}{f'}\right)^2
\end{equation*}
is the {\it Schwarzian derivative} of $f$. It is the finite form of
the infinitesimal cocycle $s$ introduced below (\ref{eq:11}). The
cocycle condition is then equivalent to the identity
\begin{equation}
  S[f\circ g]=(S[f]\circ g)(g')^2+S[g].\label{eq:65}
\end{equation}
Note that $S[h](x)=0$ iff $h(x)=(ax+b)/(cx+d)$, with $ad
-bc=1$. When the coordinate $x$ describes the projective line,
such transformations form a group isomorphic to $\PSL(2,\RR)$;
however, when $x=\phii$ is a coordinate on the circle, the only
transformation of this form that is also a diffeomorphism
of the circle is $h(\phii)=\phii+\text{cst}$.

Consider the function $t_n(\phii)=\tan(n\phii/2)$ related to the
stereographic projection of the $n$-fold cover of the circle. From
$S[t_n]=n^2/2$ and eq. \eqref{eq:65}, it follows that $S\left[t_n\circ
  f \circ t_n^{-1}\right]=(\bar S_n[f]\circ t_n^{-1})((t_n^{-1})')^2$,
with the definition
\begin{equation*}
  \bar{S}_n[f]:=S[f]+\frac{n^2}{2}((f')^2-1).
\end{equation*}
An important property of the Schwarzian derivative is the
inequality 
\begin{equation}
  \label{eq:64}
  \int_0^{2\pi} d\phii\, \bar S_1[f]\leq 0\quad\forall f\in
  \widetilde{\mathrm{Diff}^+}(S^1),
\end{equation}
or, equivalently, $\int_0^{2\pi} d\phii\,
[\theta-{(\theta')^2}/{\theta}]\leq 2\pi$ in terms of the $\theta$ function in
\eqref{eq:4}. This inequality is saturated iff $f$ is given by
\begin{equation}
  e^{if(\phii)}=\frac{\alpha e^{i\phii}+\beta}{\bar\beta
    e^{i\phii}+\bar\alpha},\quad |\alpha|^2-|\beta|^2=1,\label{eq:71}
\end{equation}
that is, iff $f$ is the lift of a projective transformation of the
circle.

Finally, the coadjoint action of the Virasoro group is given by
\begin{equation}
  {\rm Ad}^*_{f^{-1}}(p,ic)=
  \left(\left[(f')^2\,p\circ f-\frac{c}{24\pi}S[f]\right]d\phii^2,ic\right).
  \label{coadVir}
\end{equation}
The associated differential is \eqref{eq:17}, up to an overall minus
sign.

\section{BMS group in three dimensions}
\label{sec:bms3-group}

We now provide the abstract definition of $\BMS$ in terms of functions on
the circle, before passing to its more standard description on null
infinity and isolating its Poincar\'e subgroups.

\subsection{Structure of the group and algebra}
\label{sec:structure-group}

The (centerless) $\BMS$ group is the symmetry group of
three-dimensional asymptotically flat spacetimes at (future or past)
null infinity. It is the semi-direct product of {\it superrotations}
and {\it supertranslations} under the adjoint action:
\begin{equation}
  \label{eq:20}
  \BMS:=\widetilde{\mathrm{Diff}^+}(S^1)\ltimes_{\Ad}\Vect_{\rm ab}.
\end{equation}
The group of supertranslations is the abelian additive group
$\Vect_{\rm ab}$, while the group of superrotations\footnote{The
  reason for using the universal cover of the diffeomorphism group is
  that we are ultimately interested in projective representations, so
  we want our group of interest to be simply connected.}  is
$\widetilde{\mathrm{Diff}^+}(S^1)$. Elements of $\BMS$ are denoted by
$(f,\alpha)$, the adjoint action being (\ref{eq:17a}) with
$\lambda=-1$. The Lie algebra of $\BMS$ is the semi-direct sum
\begin{equation}
  \bms=\Vect\oright_{\ad}\Vect_{\rm ab}.\label{eq:39}
\end{equation}
Introducing a Fourier basis of the latter by defining
\begin{equation}
\label{bmsFourier}
j_m:=\big(e^{im\phii}\dover{}{\phii},0\big) \quad\text{and}\quad
p_m:=(0,e^{im\phii}d\phii^{-1}),
\end{equation}
the commutation relations of $\bms$ are of the form (\ref{eq:6}),
except that now the indices $m,n$ run over all integer values. This
makes explicit the fact that the $\mathfrak{bms}_3$ algebra is an
infinite-dimensional extension of the Poincar\'e algebra in three
dimensions.

As regards projective representations, the centrally extended
$\BMS$ group is relevant:
\begin{equation}
  \label{eq:21}
 \hatBMS=\hatDiff\ltimes_{\Ad}\hatVect_{\rm ab}.
\end{equation}
Its elements are quadruples $(f,-ia;\alpha,-ib)$ and the adjoint
action reads 
\begin{equation*}
\Ad_{f}(\alpha,-ib)=\left(\left(\alpha f'\right)\circ f^{-1},
\frac{i}{24\pi}\int^{2\pi}_0 d\phii\,S[f]\alpha\, -ib\right).
\end{equation*}
Up to an overall minus sign in the Lie brackets, the
associated Lie algebra is
\begin{equation}
  \label{eq:25}
  \hatbms=\hatVect\oright_{\ad}\hatVect_{\rm ab}, 
\end{equation}
whose elements are quadruples $(X,-ia;\alpha,-ib)$ with commutation
relations
\begin{multline*}
  \big[\left(X,-ia;\alpha,-ib\right),\left(Y,-ir;\beta,-is\right)\big]=\\=\left([X,Y],-
  \frac{i}{48\pi} C(X,Y);[X,\beta]-[Y,\alpha],-
  \frac{i}{48\pi} (C(X,\beta)-C(Y,\alpha))\right). 
\end{multline*}
In terms of the Fourier basis
\begin{equation}
  \label{eq:36}
\begin{split}
  J_m:=\left(e^{im\phii}\dover{}{\phii},\frac{-i}{24}\delta^0_{m};0,0\right),\quad
  Z_1:=(0,1;0,0),\\
  P_m:=\left(0,0;e^{im\phii}d\phii^{-1},\frac{-i}{24}\delta^0_{m}\right),\quad
  Z_2:=(0,0;0,1), 
\end{split}
\end{equation}
the non-vanishing brackets of $\hatbms$ are
\begin{equation}
  \begin{split}
    i[J_m,J_n]= (m-n)J_{m+n} + \frac{Z_1}{12}m(m^2-1)\delta^0_{m+n}, 
    \label{hatbms}\\
    i[J_m,P_n]= (m-n)P_{m+n}+ \frac{Z_2}{12}m(m^2-1)\delta^0_{m+n}.
  \end{split}
\end{equation}

\subsection{BMS$_3$ as transformation group on $\scri$}
\label{sec:bms3-as-transf}

So far, $\BMS$ has been given a purely one-dimensional description. In
the study of three-dimensional asymptotically flat spacetimes, an
equivalent two-dimensional description, in terms of transformations of
(future or past) null infinity $\scri =S^1\times\RR$, appears
naturally. Introducing local coordinates $(\phii,u)$ on $\scri$, the
transformation associated with $(f,\alpha)\in\BMS$ is given by
\begin{equation*}
\label{trBMS}
(\phii,u)\mapsto \lp f\big(\phii\rp,
f'(\phii)\lp u+ \alpha\lp\phii\rp\rp\big).
\end{equation*}

\subsection{Poincar\'{e} subgroups of BMS$_3$}
\label{sec:poincare-group-as}

In order to isolate the natural Poincar\'{e} subgroup of $\BMS$, it
suffices to compute the action of the Poincar\'{e} group on
$\scri$. An easy way to do this on $\scri^+$ is to use BMS coordinates
$r$, $u$, $\phii$ related to the standard cartesian coordinates
$x^\mu$ of Minkowski spacetime through $re^{i\phii}:=x^1+ix^2$,
$u:=x^0-r$. Then $r\in\RR^+$ is a radial coordinate, $u\in\RR$ is an
outgoing null coordinate, and, as before, $\phii$ is an angular
coordinate on $S^1$. In terms of BMS coordinates, the Minkowski metric
reads
\begin{equation*}
d \bar s^2=-du^2-2dudr+r^2d\phii^2.
\end{equation*} 
Poincar\'{e} transformations
$x^{\mu}\mapsto\Lambda^{\mu}_{\;\;\nu}x^{\nu}+a^{\mu}$ can then be
expressed in BMS coordinates, and the limit $r\rightarrow+\infty$
yields their action on $\scri^+$.

For a translation $x^{\mu}\mapsto x^{\mu}+a^{\mu}$, the limit
$r\rightarrow+\infty$ simply produces
\begin{equation}
\label{tran}
\lp\phii,u\rp{\mapsto}\lp\phii,
u+\alpha\rp,\quad \alpha(\phii)= a^0-a^1\cos\phii-a^2\sin\phii. 
\end{equation}
For a Lorentz transformation, it is useful to refer to the
homomorphism $H$ in (\ref{Phi}), parametrizing Lorentz transformations
by $\SL\lp2,\RR\rp$ matrices. Defining the complex variables
\begin{equation}
  \label{eq:27}
   \alpha:=\half(a+d+i(b-c)),\quad \beta:=\half(a-d-i(b+c)) \quad
   \text{such that}\;\, |\alpha|^2-|\beta|^2=1, 
\end{equation}
a Lorentz transformation is realized on $\scri^+$ through the transformation
$(\phii,u)\mapsto(\tilde\phii,\tilde u)$, where
\begin{equation}
\label{trL}
e^{i\tilde\phii}=\frac{\alpha e^{i\phii} + \beta}{\bar\beta
  e^{i\phii}+\bar\alpha},\quad \begin{pmatrix}
\alpha & \beta \\ \bar\beta & \bar\alpha
\end{pmatrix}\in \mathrm{SU}\lp 1,1\rp, 
\end{equation}
and
\begin{equation}
  \tilde u=\frac{u}{(\alpha
    e^{i\phii}+\beta) (\bar\alpha
    e^{-i\phii}+\bar\beta)}={\tilde\phii}\,'\,u.\label{eq:29a}
\end{equation}
One verifies that this action correctly reproduces the
$\SL\lp2,\RR\rp$ group law. Note that the explicit form of the
homomorphism (\ref{Phi}) in terms of complex variables is
\begin{equation*}
H\left[\begin{pmatrix}
a & b \\ c & d
\end{pmatrix}\right]=
\begin{pmatrix}
  \alpha\bar\alpha+\beta\bar\beta & \alpha\bar\beta+\bar\alpha\beta &
  i(\alpha\bar\beta-\bar\alpha\beta)\\ 
  \alpha\beta+\widebar{\alpha\beta} &
  \half(\alpha^2+\beta^2+\bar\alpha^2+\bar\beta^2) &
  \frac{i}{2}(\alpha^2-\beta^2-\bar\alpha^2+\bar\beta^2)\\ 
  i(\widebar{\alpha\beta}-\alpha\beta) &
  \frac{i}{2}(\bar\alpha^2+\bar\beta^2-\alpha^2-\beta^2)
  &\half(\alpha^2-\beta^2+\bar\alpha^2-\bar\beta^2)
\end{pmatrix}.
\end{equation*}

The natural Poincar\'e subgroup of $\BMS$ thus consists of pairs
$(f,\alpha)$ whose $f$ is of the form $f(\phii)=\tilde{\phii}(\phii)$
given by (\ref{trL}), and whose $\alpha$ is a translation of the form
(\ref{tran}). This embedding also justifies the terminology of
``superrotations'' and ``supertranslations'' introduced above.

Observe that $\BMS$ actually contains infinitely many distinct Poincar\'e
subgroups obtained by taking the $n$-fold covers of $S^1$ (with
$n\in\NN^*$). These subgroups act on the cylinder $(\phii,u)$ at null
infinity through the transformations obtained upon replacing
$\tilde\phii$ and $\phii$ by $n\tilde\phii$ and $n\phii$ in equations
(\ref{tran}), (\ref{trL}) and (\ref{eq:29a}). The associated
Poincar\'e subalgebras of $\bms$ are generated by the elements $p_0$,
$p_n$, $p_{-n}$, $j_0$, $j_n$ and $j_{-n}$ of the Fourier basis
(\ref{bmsFourier}).

\section{Induced representations of BMS$_3$}
\label{sec:induc-repr-bms3}

This section is devoted to the discussion of $\BMS$ particles, that
is, unitary irreducible representations of $\hatBMS$ obtained by the
inducing construction. As follows from the structure of this group,
the orbits that are involved are the coadjoint orbits of the Virasoro
group. We therefore revisit their classification and comment on their
physical meaning in the present context by taking inspiration from the
Poincar\'e group.

\subsection{Little algebras}
\label{sec:little-algebra}

In the $\BMS$ context, elements of $\hatVect^*_{\rm ab}$ are pairs
$(p,ic_2)$, where $c_2$ is the central charge related to $Z_2$ in
\eqref{hatbms}. The little algebra $\mathfrak g_{p,c_2}$ then consists
of vector fields $X\in\Vect$ such that $\ad^*_X(p,ic_2)=0$, or
explicitly, using eq. (\ref{eq:17}),
\begin{equation}
  \label{eq:52}
  Xp'+2X'p-\frac{c_2}{24\pi}X'''=0.
\end{equation}
The solutions $X$ of this equation depend on the central charge $c_2$ and on
the form of $p(\phii)$.

\begin{itemize}

\item For vanishing central charge $c_2$, denote by $Z$ the set of zeros
  of $p(\phii)$; there are three families of solutions:
  \begin{enumerate}
  \item If $Z=\emptyset$, the little algebra is one-dimensional.
  
  \item If $Z\neq \emptyset$ and ${\rm int}\,Z=\emptyset$, the little
    algebra is trivial.
  
  \item If $Z\neq \emptyset$ and ${\rm int}\,Z\neq\emptyset$, the
    little algebra consists of vector fields whose support is
    contained in $Z$.
  \end{enumerate}

\item For non-zero central charge $c_2$, two cases must be distinguished:
  \begin{enumerate}
  \item If $p$ belongs to the orbit of a constant $p(\phii)=k\in\RR$,
    two qualitatively different situations may occur:
    \begin{itemize}
    \item for $k=-n^2c_2/48\pi$ with $n\in \NN^*$, the
  little algebra is three-dimensional and is generated by the vector fields $\left\{
    \dover{}{\phii},\sin{n\phii}\dover{}{\phii},
    \cos{n\phii}\dover{}{\phii}\right\}$, so that $\mathfrak g_{k,c_2}\simeq
  \mathfrak{sl}^{(n)}(2,\RR)$;
    \item for $k\neq-n^2c_2/48\pi$ with $n\in \NN^*$,
      the little algebra is one-dimensional, $\mathfrak
      g_{k,c_2}\simeq\mathfrak{u}(1)$, and is generated by
      $\dover{}{\phii}$.
    \end{itemize}

  \item If $p$ does not belong to the orbit of a constant, there are
    again two families of solutions: 
\begin{itemize}
\item  either $\mathfrak g_{p,c_2}$ is a one-dimensional subalgebra of
    $\Vect$ with solutions $X$ having finitely many simple zeros;
  \item or $\mathfrak g_{p,c_2}$ is a one-dimensional subalgebra of $\Vect$
    with solutions $X$ having finitely many double zeros.
  \end{itemize}
\end{enumerate}
  
\end{itemize}

\subsection{Orbits and little groups}
\label{sec:little-groups-orbits}

The little group $G_p$ of $p$ (at fixed central charge $c_2$) consists
of all diffeomorphisms $f$ that leave $p$ fixed:
\begin{equation}
   p(f(\phii))f'(\phii)^2-\frac{c_2}{24\pi} S[f](\phii)=p(\phii),\label{eq:55}
\end{equation}
in accordance with (\ref{coadVir}). The corresponding orbit $\cO_p$ is diffeomorphic to
the following coset spaces\footnote{The notation without tilde's
  refers to the definition of $\hatBMS$ in terms of the diffeomorphism
  group instead of its universal cover.}:
  \begin{equation*}
    \cO_p\simeq\widetilde{\mathrm{Diff}^+}(S^1)/\widetilde{G}_p\simeq\Diff/G_p,
  \end{equation*}
  with $\widetilde{G}_p:=\chi^{-1}(G_p)$, where $\chi$ is the
  projection defined in (\ref{eq:56a}).
  
For vanishing central charge $c_2$, the little groups are determined
by the set $Z$ of zeros of $p$:

\begin{enumerate}
\item If $Z=\emptyset$, $p$ belongs to the orbit of the constant $k$
  given by $2\pi\sqrt{|k|}=\int^{2\pi}_0d\phii |p(\phii)|^{\half}\neq
  0$, whose sign is the same as that of $p$. The corresponding little
  group is the group $\mathrm{U}(1)$ of rigid rotations.
  \item If $Z\neq \emptyset$ and ${\rm int}\,Z=\emptyset$, the little
group is a finite cyclic subgroup of $\Diff$.
  \item If $Z\neq \emptyset$ and ${\rm int}\,Z\neq\emptyset$, the
little group is an infinite-dimensional subgroup of $\Diff$.
\end{enumerate}

For non-zero central charge $c_2$, the classification is more
involved. It can be carried out through the analysis of the monodromy
matrix $M_{\psi}$ of Hill's equation for the function $\psi$, with a
potential proportional to $p(\phii)$. Here we will mostly state the
results without proof, except for a few explicit computations in the
simplest cases. A complete discussion, including proofs, can be found
for instance in \cite{Balog:1997zz,guieu2007}. We assume $c_2>0$; the
case $c_2<0$ follows by changing the sign of $p$.

\subsubsection*{Orbits with constant representatives}

There are four families of orbits that admit a constant representative
$p=k\in\RR$:
\begin{enumerate}
  \item For $k=-n^2c_2/48\pi$ with $n\in \NN^*$, the little group is
$G_{n}:=\mathrm{PSL}^{(n)}(2,\RR)$, the $n$-fold cover of
$\mathrm{PSL}(2,\RR)$. Indeed, in that case, the stationarity
condition \eqref{eq:55} for $f$ reduces to
\begin{equation}
S[f]=\frac{n^2}{2}\left(1-(f')^{2}\right).\label{eq:54}
\end{equation}
This can be written as $\bar S_n[f]=0$ and is thus equivalent to
$S[t_n\circ f\circ t_n^{-1}](x)=0$, implying that $t_n\circ f\circ
t_n^{-1}(x)=(ax+b)/(cx+d)$ with $ad-bc=1$; the function $t_n$ is the
one introduced above (\ref{eq:64}).  This yields
\begin{equation} e^{inf(\phii)} = \frac{\alpha e^{in\phii}-\beta}
{-\bar\beta e^{in\phii}+\bar\alpha},\quad
\begin{pmatrix} \alpha & -\beta \\ -\bar\beta & \bar\alpha
\end{pmatrix} \in \mathrm{SU}(1,1),\label{eq:53}
\end{equation} where $\alpha,\beta$ are given by \eqref{eq:27} in
terms of $a,b,c,d$. (The comparison with (\ref{trL}) amounts to the
transformation mapping $M\in \SL(2,\RR)$ on $\left(M^{-1}\right)^T$.)
Furthermore, the associated
function $\theta$ defined in (\ref{eq:4}) 
is given by
\begin{equation}
\theta(\phii)=\sqrt{1+4|\alpha\beta|^2}-2|\alpha\beta|\cos{(n\phii-\phii_0)},
\label{eq:58}
\end{equation}
with $\phii_0:=\arg \alpha+\arg\beta+\pi$. In this case, the monodromy
matrix $M_{\psi}$ turns out to be of the form $(-)^n\mathbb I$, where
$\mathbb{I}$ denotes the $2\times2$ identity matrix.

\item For $k<0$ but $k\neq-n^2c_2/48\pi$ with $n\in\NN^*$, the
  stationarity condition still gives rise to equations \eqref{eq:54},
  \eqref{eq:53} and (\ref{eq:58}), now with $n$ replaced by
  $\sqrt{-48\pi k/c_2}$. In this case, the $\theta$ function
  (\ref{eq:58}) is $2\pi$-periodic iff $\alpha\beta=0$, implying
  $\beta=0$ and $f(\phii)=\phii+\text{cst}$.  The associated little
  group $G_k$ is thus isomorphic to the group $\mathrm{U}(1)$ of rigid
  rotations, whose universal cover is $\widetilde{G}_k=\RR$. These
  orbits have elliptic monodromy as the associated monodromy matrix
  turns out to satisfy $|{\rm Tr}\, M_{\psi}|>2$.
  
\item For $k>0$, equations \eqref{eq:54}, \eqref{eq:53} and
  (\ref{eq:58}) still have to hold, with $n$ replaced by $i\sqrt{48\pi
    k/c_2}$. The same reasoning as in the previous case then implies
  that the corresponding little group $G_k$ is, once more, the group
  $\mathrm{U}(1)$ of rigid rotations (with
  $\widetilde{G}_k=\RR$). These orbits have hyperbolic monodromy,
  $|{\rm Tr}\, M_\psi|<2$.

\item Finally, for $k=0$, the stationarity condition \eqref{eq:55}
reduces to $S[f]=0$, which is solved by $f=(a\phii+b)/(c\phii
+d)$. However, this defines a diffeomorphism of the circle only when
$c=0$, which implies that $f$ must be a rigid rotation. Hence the
associated little group $G_0$ is again $\mathrm{U}(1)$.  The
corresponding monodromy matrix is of parabolic type, $|{\rm Tr}\,
M_\psi|=2$.

\end{enumerate}

\subsubsection*{Orbits without constant representatives}

There are two additional families of orbits, that do not contain any
constant $p(\phii)$:
\begin{enumerate}

\item The orbits of the first family are labelled by the parameters
  $\mu>0$ and $n\in \NN^*$, and have hyperbolic monodromy. An explicit
  representative for the $n,\mu$ orbit is given by
  \begin{equation}
    \label{eq:59}
    \begin{split}
      &\frac{12\pi p(\phii)}{c_2}:=
      \mu^2+\frac{n^2+4\mu^2}{2F(\phii)}-
      \frac{3}{4} \frac{n^2}{F^2(\phii)},\\
      & F(\phii):=\cos^2{\frac{n\phii}{2}}+\left(\sin\frac{n\phii}{2}
        +\frac{2\mu}{n}\cos\frac{n\phii}{2}\right)^2>0.
    \end{split}
  \end{equation}
  Each such orbit can be understood as a tachyonic deformation of the
  orbit of the constant $-n^2c_2/48\pi$. The latter is indeed
  recovered by taking the limit $\mu\rightarrow0$ of expression
  (\ref{eq:59}). The little group $G_{n,\mu}$ associated with an
  $n,\mu$ orbit is isomorphic to $\RR^*_+\times\ZZ_n$, where $\RR^*_+$
  is the multiplicative group of positive real numbers and $\ZZ_n$ is
  the cyclic group of rigid rotations on $S^1$ by multiples of the
  angle $2\pi/n$. Its universal cover is
  $\widetilde{G}_{n,\mu}=\RR\times\ZZ$. The little algebra is
  generated by the vector field $X(\phii)\dover{}{\phii}$, with
  \begin{equation*}
  X(\phii)=\frac{1}{F(\phii)}
    \cos\frac{n\phii}{2}\left(\frac{2\mu}{n}\cos\frac{n\phii}{2}
    +\frac{2}{n}\sin\frac{n\phii}{2}\right), 
  \end{equation*}
  which has $2n$ simple zeros. 

\item The orbits of the second family are labelled by $n\in \NN^*$ and
  $\epsilon\in \{\pm 1\}$ and have parabolic monodromy. An explicit
  representative for the $n,\epsilon$ orbit is given by
  \begin{equation*}
    \begin{split}
      & \frac{12\pi p(\phii)}{c_2}:=\frac{n^2}{2H(\phii)}
      -\frac{3n^2(1+\epsilon/2\pi)}{4H^2(\phii)},\\
      & H(\phii):=1+\frac{\epsilon}{2\pi}\sin^2(n\phii/2)>0.
    \end{split}
  \end{equation*}
  The corresponding little groups are, again, $G_{n,\epsilon}\simeq
  \RR^*_+\times\ZZ_n$, $\widetilde{G}_{n,\epsilon}=\RR\times\ZZ$. Its
  Lie algebra is generated by the vector field
  $X(\phii)\dover{}{\phii}$ with
  \begin{equation*}
    X(\phii)=\frac{1}{H(\phii)}\sin^2(n\phii/2), 
  \end{equation*}
  having $n$ double zeros. Orbits of this type can be understood as
  massless deformations of constant orbits of the type $-n^2c_2/48\pi$.
\end{enumerate}

\subsection{Energy bounds}
\label{sec:energy-bounds}

The energy of a Virasoro coadjoint vector $(p,ic_2)$ is defined as 
\begin{equation}
  \label{eq:70}
  E_p:=\int^{2\pi}_0d\phii\, p(\phii).
\end{equation}
Similarly, according to (\ref{coadVir}), the energy of an element
$\Ad^*_{f^{-1}}(p,ic_2)$ belonging to the orbit $\calO_p$ of
$(p,ic_2)$ is
\begin{equation}
  \label{eq:3}
  E_p[f]=\int_{0}^{2\pi}{d\phii}\, \left[(f')^2\,p\circ
    f-\frac{c}{24\pi}S[f]\right]=
  \int_{0}^{2\pi}{d\phii}
  \left[\theta\, p+\frac{c_2}{48\pi} \frac{(\theta')^2}{\theta}\right],
\end{equation}
with $\theta$ as in \eqref{eq:4}. We will discuss the behaviour of
energy only for $c_2>0$, because the discussion for $c_2<0$ follows
from the analysis below when exchanging the words ``below'' and
``above''. (We will not consider the case $c_2=0$.) Since the last
term of (\ref{eq:3}) can be made arbitrarily large by tuning $f$, it
is obvious that the energy is unbounded from above on every orbit. But
the real question is boundedness from below.

The inequality \eqref{eq:64} readily implies that energy is
bounded from below on the orbit of the constant
$p(\phii)=-c_2/48\pi$:
\begin{equation*}
E_{-c_2/48\pi}[f]\geq-c_2/24,
\end{equation*}
with equality iff $f$ is a projective transformation of the circle as
in \eqref{eq:71}. It also follows from the property stated after
(\ref{eq:64}) that the global minimum of energy is reached precisely
at $p=-c_2/48\pi$. 

Now, for an arbitrary constant $p(\phii)=k\in\RR$, 
\begin{eqnarray*}
  E_k[f]& = &E_{-c_2/48\pi}\left[f\right]+\left(k+\frac{c_2}{48\pi}\right)
  \int_0^{2\pi}{d\phii}\, \theta\\
  & = & \left(E_{-c_2/48\pi}\left[f\right]+\frac{c_2}{24}\right) +2\pi k +
  \left(k+\frac{c_2}{48\pi}\right)\int_0^{2\pi}{d\phii}\, (f'-1)^2. 
\end{eqnarray*}
Thus, provided $k>-c_2/48\pi$, the energy on the orbit $\calO_k$ is
bounded from below:
\begin{equation*}
  E_k[f]\geq 2\pi k,
\end{equation*}
with equality iff $f=\phii+\text{cst}$, the global minimum being
reached at $p(\phii)=k$. To the contrary, if $k<-c_2/48\pi$, energy is
{\it unbounded} from below. To see this, use a boost of rapidity
$\gamma$ in the $x^1$ direction, realised in $\BMS$ through the
diffeomorphism $f_{\gamma}$ given by
\begin{equation*}
  e^{if_{\gamma}(\phii)}=\frac{\cosh(\gamma/2)e^{i\phii}+\sinh(\gamma/2)}
  {\sinh(\gamma/2)e^{i\phii}
    +\cosh(\gamma/2)}.
\end{equation*}
Such a boost preserves the energy $-c_2/24$ of $-c_2/48\pi$,
so it transforms the energy of $p=k$ into
\begin{equation*}
  -\frac{c_2}{24}+\left(2\pi k+\frac{c_2}{24}\right)\cosh\gamma.
\end{equation*}
When $k<-c_2/48\pi$, this can be made arbitrarily negative for
sufficiently large rapidity. In conclusion, on the orbit of a constant
$p(\phii)$, energy is bounded from below iff the constant
representative of the orbit is situated above $-c_2/48\pi$.

As regards orbits without constant representatives, one can show that
the energy is {\it unbounded} from below on all of them, except for the
orbit of the massless deformation of $p=-c_2/48\pi$ with
$\epsilon=-1$. In the latter case, the lower bound of
energy is $-c_2/24$, but it is not reached on the orbit.

\subsection{Features of $\BMS$ particles}
\label{sec:interpretation}

\subsubsection*{Supermomentum}
\label{sec:supermomentum}

The coadjoint vectors $(p,ic_2)$ of $\hatVect^*_{\rm ab}$ are paired
with centrally extended supertranslations $(\alpha,-ib)$ according to
formula (\ref{eq:18}) with $X$ and $a$ replaced by $\alpha$ and
$b$. The tensor density $p$ should thus be interpreted as the {\it
  supermomentum} of a $\BMS$ particle since its Fourier modes are
related to the supertranslation generators $P_m$ defined in
(\ref{eq:36}),
\begin{equation*}
  \langle(p,ic_2),P_m\rangle=
\int_0^{2\pi}d\phii\,p(\phii)e^{im\phii}+\frac{c_2}{24}\delta_{m,0}. 
\end{equation*}
In particular, the energy \eqref{eq:70} of a particle with
supermomentum $p$ is associated with the generator $P_0$ of time
translations, up to a constant shift. In addition, the Fourier modes
$p_{\pm1}$ of $p(\phii)$ encode linear momentum since they are
conjugate to spatial translations.

The Virasoro coadjoint orbits listed above are thus supermomentum
orbits of an appropriate $\BMS$ particle. From this point of view, the
orbits at non-zero central charge $c_2$, on which we now focus, are
similar to those of the Poincar\'e group: the simplest orbits with a
constant non-exceptional representative are the orbits of a $\BMS$
particle at rest, with little group $\mathrm{U}(1)$ -- exactly as for
the Poincar\'e group. The sequence of exceptional orbits sitting at
constants $-n^2c_2/48\pi$ looks like an infinite series of copies of
the vacuum orbit in the Poincar\'e case, as they all have little group
$\PSL(2,\RR)$ or a cover thereof. Finally, each of these ``vacuum
orbits'' labelled by $n$ is accompanied by two massless orbits and a
family of tachyonic orbits, which we thus take to be the orbits of
massless and tachyonic $\BMS$ particles.

Restricting physical $\BMS$ particles by the criterion of boundedness of
energy from below then requires these particles to have energy larger
than that of the only stable vacuum orbit, the one sitting at
$p=-c_2/48\pi$. The only remaining physical particles then are
1. massless particles with $\epsilon=-1$ and 2. massive particles
above $-c_2/48\pi$, the nature of which changes when the rest energy
of the particle crosses $2\pi k=0$. This qualitative change is reflected in
the change of the conjugacy class of the monodromy matrix classifying
these orbits.

\subsubsection*{BMS$_3$ spin}

With our definition (\ref{eq:20}), the $\hatBMS$ group is simply
connected. Then the appropriate little groups are the covers
$\widetilde{G}_p$ rather than the $G_p$'s. In particular, we see from
the list above that the unitary irreducible representations of each
$\widetilde{G}_p$ involve a continuous label.

Hence, if the spin of a $\BMS$ particle is defined as the label of the
chosen unitary irreducible representation of $\widetilde{G}_p$, the
spin of both massive and massless particles is {\it not} quantized,
just as in the three-dimensional Poincar\'e case. Had we defined the
$\BMS$ group in terms of $\Diff$ instead of its universal cover, this
would correspond to projective representations.

\subsubsection*{Further comments on induced representations of
  $\hatBMS$}

In order to construct a {\it unitary} induced representation, a
Virasoro-invariant measure $\calD\mu$ on coadjoint orbits is
required. The most natural way to obtain such a measure is to give a
meaning to the coadjoint symplectic form $\omega$ taken to an
``infinite power'', yielding a volume form on the corresponding orbit.
It was argued in \cite{Dai2003} that such a construction makes
sense. It is not clear to us whether this measure can indeed be used
in the present context. If yes, the rest of the construction should go
through: for the orbits of physical massive particles for instance,
the little group is abelian and one-dimensional, and is therefore
labelled by one real number $j$. In that case, the space $\calE$ of
the representation of the little group is just $\CC$. Thus the space
$\calH$ of the induced representation of a massive particle is the
space of complex-valued {\it wavefunctionals}, defined on the Virasoro
coajdoint orbit of a supermomentum with rest energy $>-c_2/48\pi$,
that are square-integrable with respect to the functional measure
$\calD\mu$. The representations obtained in this way are automatically
unitary and irreducible, and act on one-particle states with definite
supermomentum according to the formula (\ref{eq:1}).

\newpage
\section{Open questions}
\label{sec:issues-questions}

There are a number of obvious open questions that should be addressed
in the future.   

\begin{enumerate}

\item The infinite-dimensional aspects of the problem need to be better
  understood, in particular questions about the existence of an
  invariant measure on the relevant Virasoro coadjoint orbits.

\item Provided the first issue can be solved, the next problem is to
  systematically study whether the inducing construction exhausts all
  the unitary irreducible representations of $\hatBMS$, along the
  lines of what has been achieved for the globally well-defined
  version of the BMS$_4$ group.

\item A second related problem is to study the associated
  representations of the $\hatbms$ Lie algebra and work out those that
  can be related through a ``flat limit'' to the highest-weight
  representations of (two copies of) the Virasoro algebra that appear
  in the AdS case.

\item Finally, a similar analysis should be applied to the local versions of
  the $\mathrm{BMS}_4$ group and algebra.

\end{enumerate}

{\bf Note added:} The Virasoro coadjoint orbits reviewed here are also
a crucial ingredient in the discussion of the coadjoint representation
of $\hatBMS$ studied in the companion paper \cite{Barnich2014}. In
particular, the discussion on energy bounds can be used to derive
positive energy theorems for three-dimensional gravity with
non-trivial asymptotics \cite{Barnich:2014zoa}, both at null infinity
for the flat background and at spatial infinity for the anti-de Sitter
background. While completing this and the companion paper, preprint
\cite{Garbarz:2014kaa} appeared, which relies on the same understanding
of the role of Virasoro coadjoint orbits for the covariant phase space
of asymptotically ${\rm AdS}_3$ gravity.

\section*{Acknowledgements}


This work is supported in part by the Fund for Scientific
Research-FNRS (Belgium), by IISN-Belgium and by ``Communaut\'e fran\c
caise de Belgique - Actions de Recherche
Concert\'ees''. B.O. gratefully acknowledges useful discussions on
related topics with M\'elanie Bertelson, C\'edric De Groote and
Pierre-Henry Lambert.


\begin{thebibliography}{10}

\bibitem{tHooft:1993gx}
G.~'t~Hooft, ``Dimensional reduction in quantum gravity,''
\href{http://www.arXiv.org/abs/gr-qc/9310026}{{\tt gr-qc/9310026}}.

\bibitem{Susskind:1994vu}
L.~Susskind, ``The world as a hologram,'' {\em J. Math. Phys.} {\bf 36} (1995)
  6377--6396,
\href{http://www.arXiv.org/abs/hep-th/9409089}{{\tt hep-th/9409089}}.

\bibitem{Maldacena:1997re}
J.~M. Maldacena, ``{The large N limit of superconformal field theories and
  supergravity},'' {\em Adv. Theor. Math. Phys.} {\bf 2} (1998) 231--252,
\href{http://www.arXiv.org/abs/hep-th/9711200}{{\tt hep-th/9711200}}.

\bibitem{Ashtekar:1984}
A.~Ashtekar and A.~Magnon, ``Asymptotically anti-de {S}itter space-times,''
  {\em Class. Quant. Grav.} {\bf 1} (1984) L39.

\bibitem{Henneaux:1985tv}
M.~Henneaux and C.~Teitelboim, ``Asymptotically anti-de {S}itter spaces,'' {\em
  Commun. Math. Phys.} {\bf 98} (1985) 391.

\bibitem{Henneaux:1985ey}
M.~Henneaux, ``Asymptotically anti-de {S}itter universes in d = 3, 4 and higher
  dimensions,'' in {\em Proceedings of the Fourth Marcel Grossmann Meeting on
  General Relativity, Rome 1985}, R.~Ruffini, ed., pp.~959--966.
\newblock Elsevier Science Publishers B.V., 1986.

\bibitem{Ashtekar:1999jx}
A.~Ashtekar and S.~Das, ``{Asymptotically Anti-de Sitter space-times: Conserved
  quantities},'' {\em Class.Quant.Grav.} {\bf 17} (2000) L17--L30,
\href{http://www.arXiv.org/abs/hep-th/9911230}{{\tt hep-th/9911230}}.

\bibitem{Brown:1986nw}
J.~D. Brown and M.~Henneaux, ``Central charges in the canonical realization of
  asymptotic symmetries: An example from three-dimensional gravity,'' {\em
  Commun. Math. Phys.} {\bf 104} (1986) 207.

\bibitem{Belavin:1984vu}
A.~A. Belavin, A.~M. Polyakov, and A.~B. Zamolodchikov, ``{Infinite conformal
  symmetry in two-dimensional quantum field theory},'' {\em Nucl. Phys.} {\bf
  B241} (1984)
333--380.

\bibitem{Bondi:1962px}
H.~Bondi, M.~G. van~der Burg, and A.~W. Metzner, ``{Gravitational waves In
  general relativity. 7. {W}aves from axi-symmetric isolated systems},'' {\em
  Proc.\ Roy.\ Soc.\ Lond. A} {\bf 269} (1962)
21.

\bibitem{Sachs1962a}
R.~K. Sachs, ``Gravitational waves in general relativity. 8. {W}aves in
  asymptotically flat space-time,'' {\em Proc.\ Roy.\ Soc.\ Lond.\ A} {\bf 270}
  (1962)
103.

\bibitem{Sachs1962}
R.~K. Sachs, ``Asymptotic symmetries in gravitational theory,'' {\em Phys.\
  Rev.} {\bf 128} (1962) 2851--2864.

\bibitem{Strominger:2013jfa}
A.~Strominger, ``{On BMS Invariance of Gravitational Scattering},''
\href{http://www.arXiv.org/abs/1312.2229}{{\tt 1312.2229}}.

\bibitem{Strominger:2013lka}
A.~Strominger, ``{Asymptotic Symmetries of Yang-Mills Theory},''
\href{http://www.arXiv.org/abs/1308.0589}{{\tt 1308.0589}}.

\bibitem{He:2014laa}
T.~He, V.~Lysov, P.~Mitra, and A.~Strominger, ``{BMS supertranslations and
  Weinberg's soft graviton theorem},''
\href{http://www.arXiv.org/abs/1401.7026}{{\tt 1401.7026}}.

\bibitem{Banks:2014iha}
T.~Banks, ``{The Super BMS Algebra, Scattering and Holography},''
\href{http://www.arXiv.org/abs/1403.3420}{{\tt 1403.3420}}.

\bibitem{Duval:2014uva}
C.~Duval, G.~Gibbons, and P.~Horvathy, ``{Conformal Carroll groups and BMS
  symmetry},''
\href{http://www.arXiv.org/abs/1402.5894}{{\tt 1402.5894}}.

\bibitem{Newman:1965fk}
E.~T. Newman, ``A possible connexion between the gravitational field and
  elementary particle physics,'' {\em Nature} {\bf 206} (05, 1965) 811--812.

\bibitem{Komar1965}
A.~Komar, ``Quantized gravitational theory and internal symmetries,'' {\em
  Phys. Rev. Lett.} {\bf 15} (Jul, 1965) 76--78.

\bibitem{McCarthy1972}
P.~J. McCarthy, ``Asymptotically flat space-times and elementary particles,''
  {\em Phys. Rev. Lett.} {\bf 29} (Sep, 1972) 817--819.

\bibitem{Barnich:2009se}
G.~Barnich and C.~Troessaert, ``{Symmetries of asymptotically flat 4
  dimensional spacetimes at null infinity revisited},'' {\em Phys.Rev.Lett.}
  {\bf 105} (2010) 111103,
\href{http://www.arXiv.org/abs/0909.2617}{{\tt 0909.2617}}.

\bibitem{Barnich:2010eb}
G.~Barnich and C.~Troessaert, ``{Aspects of the BMS/CFT correspondence},'' {\em
  JHEP} {\bf 05} (2010) 062,
\href{http://www.arXiv.org/abs/1001.1541}{{\tt 1001.1541}}.

\bibitem{Barnich:2013axa}
G.~Barnich and C.~Troessaert, ``{Comments on holographic current algebras and
  asymptotically flat four dimensional spacetimes at null infinity},'' {\em
  JHEP} {\bf 1311} (2013) 003,
\href{http://www.arXiv.org/abs/1309.0794}{{\tt 1309.0794}}.

\bibitem{Ashtekar1997}
A.~Ashtekar, J.~Bicak, and B.~G. Schmidt, ``Asymptotic structure of
  symmetry-reduced general relativity,'' {\em Phys. Rev. D} {\bf 55} (Jan,
  1997) 669--686.

\bibitem{Barnich:2006avcorr}
G.~Barnich and G.~Comp{\`e}re, ``Classical central extension for asymptotic
  symmetries at null infinity in three spacetime dimensions,'' {\em Class.
  Quant. Grav.} {\bf 24} (2007) F15,
  \href{http://www.arXiv.org/abs/gr-qc/0610130}{{\tt gr-qc/0610130}}.
Corrigendum: ibid 24 (2007) 3139.

\bibitem{Coussaert:1995zp}
O.~Coussaert, M.~Henneaux, and P.~van Driel, ``{The asymptotic dynamics of
  three-dimensional Einstein gravity with a negative cosmological constant},''
  {\em Class. Quant. Grav.} {\bf 12} (1995) 2961--2966,
  \href{http://www.arXiv.org/abs/gr-qc/9506019}{{\tt gr-qc/9506019}}.

\bibitem{Strominger:1998eq}
A.~Strominger, ``Black hole entropy from near-horizon microstates,'' {\em JHEP}
  {\bf 02} (1998) 009,
\href{http://www.arXiv.org/abs/arXiv:hep-th/9712251}{{\tt
  arXiv:hep-th/9712251}}.

\bibitem{Banados1999}
M.~{Banados}, ``{Three-Dimensional Quantum Geometry and Black Holes},'' in {\em
  Trends in Theoretical Physics II}, vol.~484 of {\em American Institute of
  Physics Conference Series}, pp.~147--169.
\newblock 1999.
\newblock
\href{http://www.arXiv.org/abs/hep-th/9901148}{{\tt hep-th/9901148}}.
\newblock

\bibitem{Skenderis:1999nb}
K.~Skenderis and S.~N. Solodukhin, ``{Quantum effective action from the AdS/CFT
  correspondence},'' {\em Phys. Lett.} {\bf B472} (2000) 316--322,
\href{http://www.arXiv.org/abs/hep-th/9910023}{{\tt hep-th/9910023}}.

\bibitem{Barnich:2011ct}
G.~Barnich and C.~Troessaert, ``{Supertranslations call for superrotations},''
  {\em {PoS}} {\bf CNCFG2010} (2010) 010,
\href{http://www.arXiv.org/abs/1102.4632}{{\tt 1102.4632}}.

\bibitem{Barnich:2012aw}
G.~Barnich, A.~Gomberoff, and H.~A. Gonz\'alez, ``{Flat limit of three
  dimensional asymptotically anti-de Sitter spacetimes},'' {\em Phys. Rev. D}
  {\bf 86} (2012) 024020,
\href{http://www.arXiv.org/abs/1204.3288}{{\tt 1204.3288}}.

\bibitem{Barnich:2012xq}
G.~Barnich, ``{Entropy of three-dimensional asymptotically flat cosmological
  solutions},'' {\em JHEP} {\bf 1210} (2012) 095,
\href{http://www.arXiv.org/abs/1208.4371}{{\tt 1208.4371}}.

\bibitem{Bagchi:2012xr}
A.~Bagchi, S.~Detournay, R.~Fareghbal, and J.~Simon, ``{Holography of 3d Flat
  Cosmological Horizons},'' {\em Phys.Rev.Lett.} {\bf 110} (2013) 141302,
\href{http://www.arXiv.org/abs/1208.4372}{{\tt 1208.4372}}.

\bibitem{Barnich:2012rz}
G.~Barnich, A.~Gomberoff, and H.~A. Gonzalez, ``{BMS3 invariant two dimensional
  field theories as flat limit of Liouville},'' {\em Phys. Rev.} {\bf
  D87:124032,} (2013)
\href{http://www.arXiv.org/abs/1210.0731}{{\tt 1210.0731}}.

\bibitem{Barnich:2013yka}
G.~Barnich and H.~Gonzalez, ``{Dual dynamics of three dimensional
  asymptotically flat Einstein gravity at null infinity},'' {\em JHEP} {\bf
  1305} (2013) 016,
\href{http://www.arXiv.org/abs/1303.1075}{{\tt 1303.1075}}.

\bibitem{Mackey1949}
G.~W. Mackey, ``{Imprimitivity for Representations of Locally Compact Groups
  I},'' {\em Proc. Natl. Acad. Sci. U.S.A.} {\bf 35} (1949) 537--545.

\bibitem{Mackey1952}
G.~W. Mackey, ``{Induced representations of locally compact groups. I},'' {\em
  Ann. of Math.} {\bf 55} (1952) 101--139.

\bibitem{Mackey1953}
G.~W. Mackey, ``{Induced representations of locally compact groups. II. The
  Frobenius reciprocity theorem},'' {\em Ann. of Math.} {\bf 58} (1953)
  193--221.

\bibitem{MackeyBook}
G.~Mackey, {\em Induced representations of groups and quantum mechanics}.
\newblock Publicazioni della Classe di Scienze della Scuola Normale Superiore
  di Pisa. W. A. Benjamin, 1968.

\bibitem{A.O.Barut702}
A.~O. Barut and R.~Raczka, {\em Theory of Group Representations and
  Applications}.
\newblock Polish Scientific Publishers, Warszawa, 1980.

\bibitem{Cornwell:1985xs}
J.~F. Cornwell, ``Group theory in physics. vol. 1,''. London, Uk: Academic (
  1984) 399 P. ( Techniques Of Physics, 7).

\bibitem{Cornwell:1985xt}
J.~F. Cornwell, ``Group theory in physics. vol. 2,''. London, Uk: Academic (
  1984) 589 P. ( Techniques Of Physics, 7).

\bibitem{Wigner:1939cj}
E.~P. Wigner, ``{On Unitary Representations of the Inhomogeneous Lorentz
  Group},'' {\em Annals Math.} {\bf 40} (1939)
149--204.

\bibitem{Geroch:1971mz}
R.~P. Geroch and E.~Newman, ``{Application of the semidirect product of
  groups},'' {\em J.Math.Phys.} {\bf 12} (1971)
314.

\bibitem{McCarthy1972b}
P.~J. McCarthy, ``{Structure of the Bondi-Metzner-Sachs Group},'' {\em Journal
  of Mathematical Physics} {\bf 13} (1972), no.~11, 1837--1842.

\bibitem{McCarthy1972a}
P.~J. {McCarthy}, ``{Representations of the Bondi-Metzner-Sachs Group. I.
  Determination of the Representations},'' {\em Royal Society of London
  Proceedings Series A} {\bf 330} (Nov., 1972) 517--535.

\bibitem{McCarthy1973}
P.~J. {McCarthy}, ``{Representations of the Bondi-Metzner-Sachs Group. II.
  Properties and Classification of the Representations},'' {\em Royal Society
  of London Proceedings Series A} {\bf 333} (May, 1973) 317--336.

\bibitem{McCarthy1973a}
P.~J. {McCarthy} and M.~{Crampin}, ``{Representations of the
  Bondi-Metzner-Sachs Group. III. Poincare Spin Multiplicities and
  Irreducibility},'' {\em Royal Society of London Proceedings Series A} {\bf
  335} (Nov., 1973) 301--311.

\bibitem{McCarthy1975}
P.~J. {McCarthy}, ``{The Bondi-Metzner-Sachs Group in the Nuclear Topology},''
  {\em Royal Society of London Proceedings Series A} {\bf 343} (May, 1975)
  489--523.

\bibitem{McCarthy1978}
P.~J. {McCarthy}, ``{Hyperfunctions and Asymptotic Symmetries},'' {\em Royal
  Society of London Proceedings Series A} {\bf 358} (Jan., 1978) 495--498.

\bibitem{McCarthy1978a}
P.~McCarthy, ``Lifting of projective representations of the
  {B}ondi-{M}etzner-{S}achs group,'' {\em Proc. R. Soc. London} {\bf A 358}
  (1978) 141--171.

\bibitem{Piard1977}
A.~Piard, ``Unitary representations of semi-direct product groups with infinite
  dimensional abelian normal subgroup,'' {\em Reports on Mathematical Physics}
  {\bf 11} (1977), no.~2, 259 -- 278.

\bibitem{Piard1977a}
A.~Piard, ``{Representations of the Bondi-Metzner-Sachs group with the Hilbert
  topology},'' {\em Reports on Mathematical Physics} {\bf 11} (1977), no.~2,
  279 -- 283.

\bibitem{A.A.Kirillov897}
A.~A. Kirillov, {\em Lectures on the orbit method}.
\newblock American Mathematical Society, 2004.

\bibitem{LazPan}
V.~Lazutkin and T.~Pankratova, ``{Normal forms and versal deformations for
  Hill's equation},'' {\em Funkts. Anal. Prilozh.} {\bf 9} (1975) 41--48.

\bibitem{Segal:1981ap}
G.~Segal, ``{Unitarity Representations of Some Infinite Dimensional Groups},''
  {\em Commun.Math.Phys.} {\bf 80} (1981)
301--342.

\bibitem{Witten:1987ty}
E.~Witten, ``{Coadjoint Orbits of the Virasoro Group},'' {\em Commun. Math.
  Phys.} {\bf 114} (1988)
1.

\bibitem{Bakas:1988mq}
I.~Bakas, ``{Conformal Invariance, the KdV Equation and Coadjoint Orbits of the
  Virasoro Algebra},'' {\em Nucl.Phys.} {\bf B302} (1988)
189--203.

\bibitem{Taylor:1992xt}
W.~Taylor, ``{Virasoro representations on diff S1 / S1 coadjoint orbits},''
\href{http://www.arXiv.org/abs/hep-th/9204091}{{\tt hep-th/9204091}}.

\bibitem{Henkel2005}
M.~{Henkel}, R.~{Schott}, S.~{Stoimenov}, and J.~{Unterberger}, ``{On the
  dynamical symmetric algebra of ageing: Lie structure, representations and
  Appell systems},'' {\em ArXiv Mathematics e-prints} (Oct., 2005)
  \href{http://www.arXiv.org/abs/arXiv:math/0510096}{{\tt arXiv:math/0510096}}.

\bibitem{Bagchi:2009my}
A.~Bagchi and R.~Gopakumar, ``{Galilean Conformal Algebras and AdS/CFT},'' {\em
  JHEP} {\bf 07} (2009) 037,
\href{http://www.arXiv.org/abs/0902.1385}{{\tt 0902.1385}}.

\bibitem{Hosseiny:2009jj}
A.~Hosseiny and S.~Rouhani, ``{Affine Extension of Galilean Conformal Algebra
  in 2+1 Dimensions},'' {\em J.Math.Phys.} {\bf 51} (2010) 052307,
\href{http://www.arXiv.org/abs/0909.1203}{{\tt 0909.1203}}.

\bibitem{Bagchi:2009pe}
A.~Bagchi, R.~Gopakumar, I.~Mandal, and A.~Miwa, ``{GCA in 2d},'' {\em JHEP}
  {\bf 08} (2010) 004,
\href{http://www.arXiv.org/abs/0912.1090}{{\tt 0912.1090}}.

\bibitem{Bargmann1954}
V.~Bargmann, ``On unitary representations of continuous groups,'' {\em Annals
  of Mathematics} {\bf 59} (1954), no.~1, 1--46.

\bibitem{Simms1971}
D.~Simms, {\em {Lie Groups and Quantum Mechanics}}, vol.~52 of {\em Lecture
  Notes in Mathematics}.
\newblock Springer, 1968.

\bibitem{Weinberg:1995mt}
S.~Weinberg, {\em The {Q}uantum {T}heory of {F}ields. Vol. 1: Foundations}.
\newblock Cambridge University Press, 1995.

\bibitem{Binegar1982}
B.~Binegar, ``Relativistic field theories in three dimensions,'' {\em Journal
  of Mathematical Physics} {\bf 23} (1982), no.~8, 1511--1517.

\bibitem{Ovsienko2005}
V.~Ovsienko and S.~Tabachnikov, {\em {Projective Differential Geometry Old and
  New. From the Schwarzian Derivative to Cohomology of Diffeomorphism Groups}}.
\newblock Cambridge University Press, 2005.

\bibitem{guieu2007}
L.~Guieu and C.~Roger, {\em {L'Alg\`ebre et le Groupe de Virasoro}}.
\newblock {Les Publications CRM, Montr\'eal}, 2007.

\bibitem{Khesin2009}
B.~Khesin and R.~Wendt, {\em {The Geometry of Infinite-Dimensional Groups}}.
\newblock Springer-Verlag Berlin Heidelberg, 2009.

\bibitem{Unterberger:2011yya}
J.~Unterberger and C.~Roger, {\em {The Schr\"odinger-Virasoro Algebra:
  Mathematical structure and dynamical Schr\"odinger symmetries}}.
\newblock Springer,
2012.
\newblock

\bibitem{Balog:1997zz}
J.~Balog, L.~Feher, and L.~Palla, ``{Coadjoint orbits of the Virasoro algebra
  and the global Liouville equation},'' {\em Int. J. Mod. Phys.} {\bf A13}
  (1998) 315--362,
\href{http://www.arXiv.org/abs/hep-th/9703045}{{\tt hep-th/9703045}}.

\bibitem{Dai2003}
J.~Dai and D.~Pickrell, ``{The orbit method and the Virasoro extension of
  Diff+(S1): I. Orbital integrals},'' {\em Journal of Geometry and Physics}
  {\bf 44} (2003), no.~4, 623 -- 653.

\bibitem{Barnich2014}
G.~Barnich and B.~Oblak, ``{Notes on the BMS group in three dimensions: II.
  Coadjoint representation}.'' to appear, 2014.

\bibitem{Barnich:2014zoa}
G.~Barnich and B.~Oblak, ``{Holographic positive energy theorems in
  three-dimensional gravity},''
\href{http://www.arXiv.org/abs/1403.3835}{{\tt 1403.3835}}.

\bibitem{Garbarz:2014kaa}
A.~Garbarz and M.~Leston, ``{Classification of Boundary Gravitons in AdS$_3$
  Gravity},''
\href{http://www.arXiv.org/abs/1403.3367}{{\tt 1403.3367}}.

\end{thebibliography}


\newpage
\section*{References}

\renewcommand{\section}[2]{}%

\def\cprime{$'$}
\providecommand{\href}[2]{#2}\begingroup\raggedright\endgroup

\end{document}